\documentclass[showpacs,preprintnumbers,amsmath,amssymb,prb]{revtex4}

\def\Xint#1{\mathchoice
   {\XXint\displaystyle\textstyle{#1}}%
   {\XXint\textstyle\scriptstyle{#1}}%
   {\XXint\scriptstyle\scriptscriptstyle{#1}}%
   {\XXint\scriptscriptstyle\scriptscriptstyle{#1}}%
   \!\int}
\def\XXint#1#2#3{{\setbox0=\hbox{$#1{#2#3}{\int}$}
     \vcenter{\hbox{$#2#3$}}\kern-.5\wd0}}

\def\dashint{\Xint-}

\usepackage{graphicx}
\usepackage{subfigure}
\usepackage{dcolumn}
\usepackage{bm}

\newcommand{\kap}{\mbox{\boldmath $\kappa$}}

\begin{document}

\title{Localized and propagating excitations in gapped phases of spin systems with bond disorder}

\author{O.\ I.\ Utesov$^1$}
\email{utiosov@gmail.com}
\author{A.\ V.\ Sizanov$^{1,2}$}
\email{alexey.sizanov@gmail.com}
\author{A.\ V.\ Syromyatnikov$^{1,2}$}
\email{asyromyatnikov@yandex.ru}
\affiliation{$^1$Petersburg Nuclear Physics Institute NRC "Kurchatov Institute", Gatchina, St.\ Petersburg 188300, Russia}
\affiliation{$^2$Department of Physics, Saint Petersburg State University, 198504 St.\ Petersburg, Russia}

\date{\today}

\begin{abstract}

Using the conventional $T$-matrix approach, we discuss gapped phases in 1D, 2D, and 3D spin systems (both with and without a long range magnetic order) with bond disorder and with weakly interacting bosonic elementary excitations. This work is motivated by recent experimental and theoretical activity in spin-liquid-like systems with disorder and in the disordered interacting boson problem. In particular, we apply our theory to both paramagnetic low-field and fully polarized high-field phases in dimerized spin-$\frac12$ systems and in integer-spin magnets with large single-ion easy-plane anisotropy $\cal D$ with disorder in exchange coupling constants (and/or $\cal D$). The elementary excitation spectrum and the density of states are calculated in the first order in defects concentration $c\ll1$. In 2D and 3D systems, the scattering on defects leads to a finite damping of all propagating excitations in the band except for states lying near its edges. We demonstrate that the analytical approach is inapplicable for states near the band edges and our numerical calculations reveal their {\it localized} nature. 
We find that the damping of propagating excitations can be much more pronounced in considered systems than in magnetically ordered gapless magnets with impurities.
In 1D systems, the disorder leads to localization of {\it all} states in the band, while those lying far from the band edges (short-wavelength excitations) can look like conventional wavepackets.

\end{abstract}

\pacs{75.10.Jm, 75.10.Kt, 75.10.Pq}

\maketitle

\section{Introduction}

Even small amount of disorder can change considerably some properties of condensed matter systems. The most famous examples are probably the Anderson localization \cite{Evers} and the Kondo effect \cite{Hewson}. Disordered boson systems (so-called dirty-boson systems) have attracted much attention recently because a possibility of studying some peculiar predictions in this field has arisen in magnetically disordered spin-liquid-like materials and in optical lattices of ultracold atoms (see Ref.~\cite{Zhel13} for review). In particular, the existence of a disordered gapless Bose-glass (BG) phase was predicted for dirty bosons between gapped Mott-insulating (MI) and gapless superfluid (SF) phases. \cite{Fisher} A general theorem has been proven recently which states that BG phase always intervenes between MI and SF phases. \cite{theorem} The transition between MI and BG phases takes place via the Griffiths mechanism. \cite{grif} The nature of the quantum phase transition from BG to SF phases has been widely debated in recent years (see Refs.~\cite{Zhel13,kisel} and references therein).

\begin{figure}
  \noindent
  \includegraphics[scale=0.8]{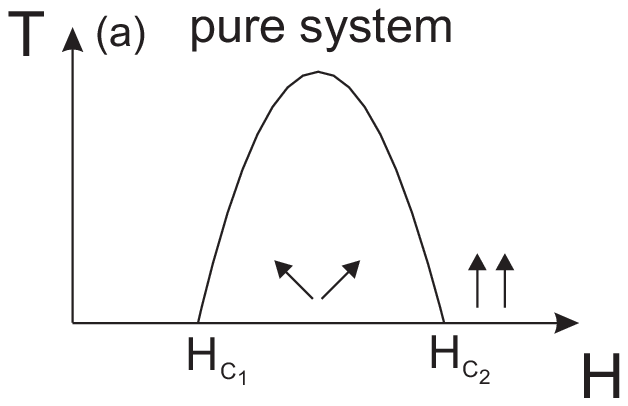}
  \includegraphics[scale=0.8]{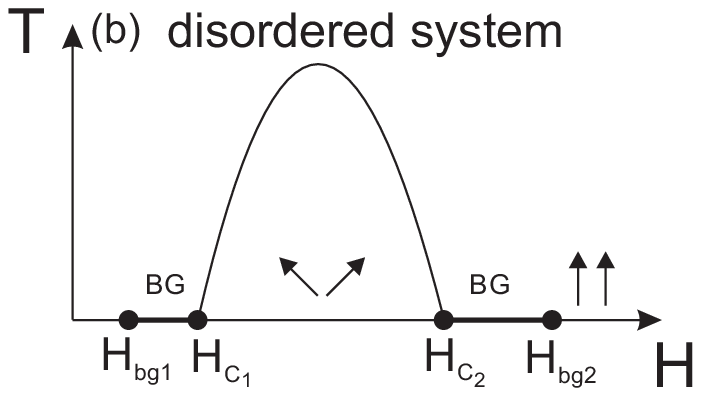}
  \hfil
  \caption{Phase diagrams of (quasi-)3D dimerized spin-$\frac12$ systems and spin-1 magnets with large single-ion easy-plane anisotropy $\cal D$ in magnetic field $H$. (a) Systems without defects with a canted magnetic ordering inside the dome. (b) Systems with a small fraction of bonds with strengthen and/or weaken exchange coupling constants (and/or $\cal D$ value). Bose-glass phases are denoted as BG.}
  \label{th}
\end{figure}

It has been understood recently that spin-1 magnets with large single-ion easy-plane anisotropy $\cal D$ and spin-$\frac12$ dimerized systems are convenient objects for discussing the dirty boson problem if disorder is realized in exchange coupling constants and/or $\cal D$. \cite{Zhel13} Such systems can be prepared in practice by creating a disorder on peripheral sites involved in superexchange interactions. A number of both large-$\cal D$ and dimerized substances with such disorder have been synthesized to date. \cite{Zhel13} At small magnetic field $H$, pure systems of this type have singlet ground states separated from the triplet excitation bands by gaps. For quasi-1D, quasi-2D and 3D systems, the phase diagram on the $T-H$ plane is presented in Fig.~\ref{th}(a) that shows a magnetically ordered gapless (SF) phase at $H_{c1}<H<H_{c2}$ and paramagnetic gapped (MI) phases at $H<H_{c1}$ and $H>H_{c2}$ (the fully polarized phase).

To explain the problem we address in the present paper, let us strengthen some randomly chosen intradimer exchange coupling constants $\cal J$ (or $\cal D$ values on some sites) in these systems. Localized impurity levels can appear inside the gap in the fully polarized phase which start to "condense" at some critical field $H_{bg2}>H_{c2}$ transferring the system into the BG phase (see Fig.~\ref{th}(b)). There are magnetically ordered islands around "strong" defects in this high-field BG phase which are well separated from each other by a nonmagnetic background, while no coherent long-range magnetic order exists in the whole system. Local quantized axes align in all islands (or all islands merge) when the transition to the magnetically ordered phase takes place at $H=H_{c2}$. \cite{foot} In contrast to the fully polarized phase, there are no localized impurity levels inside the gap for "strong" defects at small $H$. Nevertheless the general theorem \cite{theorem} requires that the field-induced transition to the ordered phase should take place via a BG phase. On the other hand, the Zeeman term commutes with the Hamiltonian and the magnetic field plays a role of a chemical potential at small and large $H$ in bosonic analogues of spin Hamiltonians (see also below for detail). Then, one is lead to a somewhat counterintuitive conclusion that at least low-energy states in the excitation band are localized at $H=0$ in the case of "strong" defects and their "condensation" drives the system into the BG phase at $H_{bg1}<H_{c1}$ (see Fig.~\ref{th}(b)). \cite{foot} Analogously, one leads to the same counterintuitive conclusion for "weak" defects at large $H$. As a result a natural question arises: which of the states in the band become localized and which of them remain propagating acquiring only a finite damping due to scattering on defects. This question looks particularly important in the light of recent excitation spectra measurements in IPA-Cu(Cl$_x$Br$_{1-x}$)$_3$ (Ref.~\cite{nafradi}) and $\rm (C_4H_{12}N_2)Cu_2(Cl$$_{1-x}$Br$_x$)$_6$ (Ref.~\cite{huv2}), dimerized materials with no impurity levels inside the gap at $H=0$. Despite considerable interest to bond disorder in spin-liquid-like magnets, this question has not been raised yet \cite{Zhel13,Vojta,Arlego} and the possibility of localization of some states in the band has not been considered in the experimental papers Refs.~\cite{nafradi,huv2}.

We attack this problem analytically using the conventional $T$-matrix approach that is widely used in discussion of defects in condensed matter theory \cite{donia} and was proven to be very useful for magnetically ordered systems with impurities. \cite{Izyumov,white,orderdif,wan,2dvac,referee1,referee2,igar} This approach allows to find corrections to Green's functions, the excitation spectrum and the density of states (DOS) in the first order in defects concentration $c\ll1$. If the expansion in terms of $c$ is valid, excitations remain propagating in disordered systems although a finite damping arises due to scattering on defects. It can happen, however, that terms of higher-order in $c$ are also important for some momenta $\bf k$ signifying analytical approach inapplicability and necessity of an additional analysis. This can be a sign of propagating modes resonance scattering on defects (which, however, can remain propagating as a result of this scattering) \cite{Izyumov,white} or a localization of some states (see, e.g., Ref.~\cite{2dvac}).

In dimerized spin-$\frac12$ systems and in integer-spin large-$\cal D$ magnets, the $T$-matrix approach allows to perform a unified consideration of all gapped phases, because Green's functions and spectra of propagating excitations have the same form. The only formal requirement to be fulfilled is that excitations in pure systems are weakly interacting. This condition holds at small $H$ if the intradimer exchange constant $\cal J$ and $\cal D$ are much larger than other exchange coupling constants $J_{ij}$. In the fully polarized phase, the magnon interaction does not lead to renormalization of observables at $T=0$ and it can be omitted. \cite{chub}  We consider below the disorder in $\cal J$ or $\cal D$ as well as in $J_{ij}$. It is found that the analytical approach is invalid in 1D, 2D and 3D systems for states near the bottom and the top of excitation bands for all kinds of bond disorder (i.e., for "strong" and "weak" defects and for systems containing both "strong" and "weak" impurities). The linear size of regions in the $\bf k$-space inside which the analytical approach does not work scale as some powers of $c$. To clarify the nature of states near band edges, we perform numerical calculations for 1D and 2D systems which show that these states are {\it localized} and they have nothing to do with conventional wavepackets. In 2D systems, the numerical analysis shows that states inside the band far from its edges are well-defined wavepackets which energies and lifetimes are given by analytical expressions obtained in the first order in $c$ (one expects the same conclusion in 3D systems). In 1D systems, {\it all} states in the band are found to be localized (similar to 1D electronic systems). At the same time, some of the states inside the band reflect properties of propagating short-wavelength excitations which energies and lifetimes are given by analytical expressions obtained in the first order in $c$. Besides, it is found that some states inside the band in 1D systems are not conventional wavepackets due to a resonant scattering on strong enough defects.


Our spectrum calculations show that in the vicinity of $H_{bg1}$ or $H_{bg2}$, if no localized impurity levels exist in the gap, the ratio of the long-wavelength propagating modes damping $\gamma_{\bf k}$ to their energy $\varepsilon_{\bf k}$ can reach $c/k^2$ in the range of this result validity $1\gg k\gg \sqrt c$. This contrasts with magnetically ordered gapless magnets in which $\gamma_{\bf k}/\varepsilon_{\bf k}$ does not exceed $c$. \cite{wan,2dvac,syromyat1,syromyat2} Thus, the damping of propagating excitations can be much more pronounced in considered systems than in magnetically ordered gapless magnets with impurities.

The results obtained can be relevant to other gapped phases in bond disordered spin systems both with and without a long range magnetic order (e.g., bond disordered easy axis ferromagnets and antiferromagnets with large easy axis anisotropy). Our main analytical results are represented in quite a model-independent form that allows using them in analysis of other systems.

The rest of the present paper is organized as follows. Pure systems are considered in Sec.~\ref{puresec}, where we derive bosonic analogs of spin Hamiltonians in all gapped phases using standard spin operators representations and demonstrate their similarity in dimerized and large-$\cal D$ systems. Our analytical and numerical methods are described in Sec.~\ref{method}. Bond disordered systems are considered in Sec.~\ref{disordersec}. Sec.~\ref{sum} contains a summary of results and our conclusions. An appendix is added with details of calculations.

\section{Gapped phases in pure systems}
\label{puresec}

In this section, we derive Bose-analogs of spin Hamiltonians describing dimerized and large-$\cal D$ systems at $H<H_{c1}$ and  $H>H_{c2}$ and demonstrate their similarity that allows the subsequent unified consideration. We derive elementary excitation spectra neglecting interaction between quasiparticles. This harmonic approximation is justified at $H>H_{c2}$ because spin-wave interaction does not modify one-particle Green's functions in the fully polarized phase. \cite{chub} At $H<H_{c1}$, the quasiparticle interaction can be neglected in the first order in the small exchange coupling $J_{ij}$ of spins from different dimers or from different sites (in large-$\cal D$ systems).

It is shown below that spectra of all modes in this approximation have the form
\begin{equation}
	\varepsilon_{\bf k} = \Delta + \frac a2 (J_{\bf k}-J_{\bf k_0}),
	\label{spec}
\end{equation}
where $a>0$ is a constant, $\Delta$ is the gap value, $J_\mathbf{k}$ is the Fourier transform of $J_{ij}$, and ${\bf k}_0$ is the momentum at which $J_{\bf k}$ reaches its minimum. For simplicity, we assume below that $J_{ij}\ne0$ for nearest neighbors only and that $J_{ij}$ either positive or negative so that all components of ${\bf k}_0$ are equal to $\pi$ if $J_{ij}>0$ and ${\bf k}_0={\bf 0}$ when $J_{ij}<0$. Then, $\varepsilon_{\bf k}$ depends quadratically on $\kap={\bf k}-{\bf k}_0$ near its minimum:
\begin{equation}
	\varepsilon_{\bf k} = \Delta + \frac a2|J| \kappa^2.
	\label{spec0}
\end{equation}
One obtains similar quadratic dependence near the spectrum maximum: $\varepsilon_{\bf k} = \Delta +a|J_{\bf k_0}|- a|J| \kappa^2/2$. Here and below $\kap$ measures a deviation of the momentum from values at which the bare spectrum has a minimum or a maximum. All the results obtained in this section are not original. We omit some details of the corresponding consideration which can be found in cited papers.

\subsection{Spin-1/2 dimer systems}

We discuss 1D, 2D and 3D simple Bravias lattices of spin-$\frac12$ dimers which Hamiltonian is written in the following form:
\begin{equation}
  \mathcal{H}=\sum_i {\cal J} {\mathbf S}_{i,1} \cdot \mathbf{S}_{i,2} + \sum_{ m} \sum_{\langle i,j\rangle}  J_{ij} \left( \mathbf{S}_{i,1} \cdot \mathbf{S}_{j,1} + \mathbf{S}_{i,2} \cdot \mathbf{S}_{j,2}\right) - h \sum_i \left(S^z_{i,1}+S^z_{i,2} \right),
  \label{ham1}
\end{equation}
where $\mathbf{S}_{i,n}$ denotes $n$-th spin ($n=1,2$) in $i$-th dimer, $h=g\mu_BH$ is the external magnetic field and $\langle i,j\rangle$ denote nearest neighbor dimers. We set below the intradimer coupling constant ${\cal J}=1$. The exchange coupling between spins from different dimers in Eq.~\eqref{ham1} is taken in the simplest form.

\subsubsection{$H<H_{c1}$}

The system has a singlet ground state that corresponds to the paramagnetic phase in Fig.~\ref{th}(a). We derive the Bose-analog of spin Hamiltonian \eqref{ham1} in the standard way \cite{sach} by introducing three Bose-operators $\mathfrak a$, $\mathfrak b$, and $\mathfrak c$ for each dimerized bond which act on the vacuum spin state $|0\rangle=\frac{1}{\sqrt{2}}\left(|\uparrow\downarrow\rangle-|\downarrow\uparrow\rangle \right)$ as follows: $\mathfrak a|0\rangle=\mathfrak b|0\rangle=\mathfrak c|0\rangle=0$, $\mathfrak a^+|0\rangle=|\uparrow\uparrow\rangle$, $\mathfrak b^+|0\rangle=|\downarrow\downarrow\rangle$, and $\mathfrak c^+|0\rangle=\frac{1}{\sqrt{2}}\left(|\uparrow\downarrow\rangle+|\downarrow\uparrow\rangle \right)$. One has for spin operators
\begin{equation}
  \begin{split}
    S^+_{i,1}=\frac{1}{\sqrt{2}}(\mathfrak a_i^+(\mathfrak c_i-1)+(\mathfrak c_i^++1)\mathfrak b_i),\\
    S^+_{i,2}=\frac{1}{\sqrt{2}}(\mathfrak a_i^+(\mathfrak c_i+1)+(\mathfrak c_i^+-1)\mathfrak b_i),\\
    S^z_{i,1}=\frac{1}{2}((\mathfrak c_i^++\mathfrak c_i)+\mathfrak a_i^+\mathfrak a_i-\mathfrak b_i^+\mathfrak b_i),\\
    S^z_{i,2}=\frac{1}{2}(-(\mathfrak c_i^++\mathfrak c_i)+\mathfrak a_i^+\mathfrak a_i-\mathfrak b_i^+\mathfrak b_i).
  \end{split}
  \label{spinrep}
\end{equation}
To fulfill the requirement that no more than one triplon $\mathfrak a$, $\mathfrak b$ or $\mathfrak c$ can sit on the same bond, one has to introduce constraint terms into the Hamiltonian which describe an infinite repulsion between triplons $U \sum_i(\mathfrak a^+_i \mathfrak a^+_i \mathfrak a_i \mathfrak a_i + \mathfrak b^+_i \mathfrak b^+_i \mathfrak b_i \mathfrak b_i + \mathfrak c^+_i \mathfrak c^+_i \mathfrak c_i \mathfrak c_i + \mathfrak a^+_i \mathfrak b^+_i \mathfrak a_i \mathfrak b_i + \mathfrak a^+_i \mathfrak c^+_i \mathfrak a_i \mathfrak c_i + \mathfrak b^+_i \mathfrak c^+_i \mathfrak b_i \mathfrak c_i)$, where $U \to +\infty$.

After substituting Eqs.~\eqref{spinrep} into Eq.~\eqref{ham1} one obtains the Bose-analog of the spin Hamiltonian which contains the constant term and terms with products of two and four Bose-operators. We restrict ourselves below by calculating triplon spectra in the first order in the interdimer coupling $J_{ij}$. It can be shown (see, e.g., Ref.~\cite{oleg}) that triplons spectra are defined only by bilinear part of the Hamiltonian in the first order in $J_{ij}$ and one has to take into account quasiparticles interaction to find spectra in higher orders. Then, the bilinear part of the Hamiltonian
\begin{equation}
 \mathcal{H}_2 = \sum_{{\bf k}} \left[ \left( 1 + \frac{J_{\bf k}}{2}-h\right)\mathfrak a^+_{\bf k} \mathfrak a_{\bf k} + \left( 1 + \frac{J_{\bf k}}{2}+h\right)\mathfrak b^+_{\bf k} \mathfrak b_{\bf k} + \left( 1 + \frac{J_{\bf k}}{2}\right)\mathfrak c^+_{\bf k} \mathfrak c_{\bf k} - \frac{J_{\bf k}}{2} (\mathfrak a_{\bf k} \mathfrak b_{-{\bf k}} + \mathfrak a^+_{\bf k} \mathfrak b^+_{-{\bf k}}) +\frac{J_{\bf k}}{4}(\mathfrak c_{\bf k} \mathfrak c_{-{\bf k}} + \mathfrak c^+_{\bf k} \mathfrak c^+_{-{\bf k}}) \right]
\label{h2}
\end{equation}
gives Eq.~\eqref{spec} for triplons spectra in the first order in $J_{ij}$ with $a=1$, $\Delta_{\mathfrak a} = 1-h+\frac12J_{{\bf k}_0}$, $\Delta_{\mathfrak b} = 1+h+\frac12J_{{\bf k}_0}$, and $\Delta_{\mathfrak c} = 1+\frac12J_{{\bf k}_0}$. As the last two terms in Eq.~\eqref{h2} do not contribute to spectra in the first order in $J_{ij}$, we omit them in the subsequent consideration.

\subsubsection{$H>H_{c2}$}

One can use the Holstein-Primakoff spin representation in the fully polarized phase at $H>H_{c2}$. As soon as magnon interaction does not lead to spectrum renormalization in this case, \cite{chub} we restrict ourselves with the linear spin-wave approximation and use the following expressions:
\begin{equation}
    S^x_{i,n}=\frac{1}{2}(\mathfrak a_{i,n}+\mathfrak a^+_{i,n}), \quad
    S^y_{i,n}=-\frac{i}{2}(\mathfrak a_{i,n}-\mathfrak a^+_{i,n}), \quad
    S^z_{i,n}=\frac{1}{2}-\mathfrak a^+_{i,n}\mathfrak a_{i,n}.
		\label{hp}
\end{equation}
After the Hamiltonian transformation and introduction of new Bose-operators
\begin{equation}
  \mathfrak a_{i,I}=\frac{\mathfrak a_{i,1}+\mathfrak a_{i,2}}{\sqrt{2}}\quad \mbox{and}\quad \mathfrak a_{i,II}=\frac{\mathfrak a_{i,1}-\mathfrak a_{i,2}}{\sqrt{2}},
\end{equation}
one obtains
\begin{equation}
  \mathcal{H}_2 = \sum_{{\bf k}} \left[ \left(h-\frac12J_{\bf 0}+\frac12J_{\bf k}\right)\mathfrak a^+_{{\bf k},I}\mathfrak a_{{\bf k},I}^{} + \left(h-1-\frac12J_{\bf 0}+\frac12J_{\bf k}\right)\mathfrak a^+_{{\bf k},II}\mathfrak a_{{\bf k},II}^{}\right],
	\label{h22}
\end{equation}
where two branches of elementary excitations have spectra of the form \eqref{spec} with $a=1$.

\subsection{Systems with integer spin and large single-ion easy-plane anisotropy}

We consider the following Hamiltonian for such systems
\begin{equation}
  \mathcal{H}=\sum_{\langle i,j\rangle} J_{ij} \mathbf{S}_i \cdot \mathbf{S}_j + {\cal D} \sum_{i} (S^z_i)^2 -  h \sum_i S^z_{i},
  \label{ham2}
\end{equation}
where ${\cal D}>0$ and ${\cal D} \gg |J_{ij}|$. Similar to spin-dimer systems, these ones have singlet (paramagnetic) ground states at small $h$ in which all spins are predominantly in states with $S^z=0$. If $S=1$, the system has the $T$--$H$ diagram shown in Fig.~\ref{th}(a). For greater integer $S$, the phase diagram contains $S$ separated (if $|J_{ij}|$ is small enough) regions with canted magnetic ordering. \cite{chub} In this case, notations $H_{c1}$ and $H_{c2}$ used below denote the smallest and the largest critical fields, respectively. Similar to the intradimer coupling constant $\cal J$, we set below ${\cal D}=1$.

\subsubsection{$H<H_{c1}$}

For arbitrary integer spin $S$, the Bose-analog of spin Hamiltonian \eqref{ham2} can be derived using the following representation (see Ref.~\cite{Sizanov} for details):
\begin{eqnarray}
  S^+_i &=& \mathfrak b^+_i (f_1-f_2 \mathfrak b^+_i \mathfrak b_i)+(f_1-f_2 \mathfrak a^+_i \mathfrak a_i)\mathfrak a_i,\nonumber\\
  S^z_i &=& \mathfrak b^+_i \mathfrak b_i -\mathfrak a^+_i \mathfrak a_i,
	\label{drep}
\end{eqnarray}
where two types of Bose-operators are introduced and
\begin{equation}
  f_1=\sqrt{S(S+1)},
	\quad
  f_2=\sqrt{S(S+1)}-\sqrt{(S-1)(S+2)/2}>0.
\end{equation}
To obtain quasiparticles spectra in the first order in the exchange interaction, one needs only the bilinear part of the Hamiltonian \cite{Sizanov} which has the form
\begin{equation}
 \mathcal{H}_2 = \sum_{{\bf k}} \left[ \left(1 + h + \frac{f^2_1}{2}J_{\bf k}\right)\mathfrak a^+_{\bf k} \mathfrak a_{\bf k} + \left(1 - h + \frac{f^2_1}{2}J_{\bf k}\right)\mathfrak b^+_{\bf k} \mathfrak b_{\bf k}\right]
  \label{h23}
\end{equation}
and which describes two branches of excitations with spectra \eqref{spec}, where $a=f^2_1=S(S+1)$.

\subsubsection{$H>H_{c2}$}

The high-field fully polarized phase can be considered using the Holstein-Primakoff spin representation that gives for the Hamiltonian in the linear spin-wave approximation
\begin{equation}
  \mathcal{H}_2 = \sum_{{\bf k}} \left(h-(2S-1)-S(J_{\bf 0}-J_{\bf k})\right)\mathfrak a^+_{\bf k} \mathfrak a_{\bf k}.
	 \label{h24}
\end{equation}
The spectrum has the form \eqref{spec} in this case with $a=2S$.

It should be noted that magnetic field can only change the gap value but does not affect the strength of the quasiparticles interaction at $T=0$ and plays the role of a chemical potential. This is related to commutation of the Zeeman term with spin Hamiltonians \eqref{ham1}, \eqref{ham2} and to particular forms of spin representations \eqref{spinrep}, \eqref{hp}, and \eqref{drep}. Magnetic field reduces the gap in the spectrum of one of the branches, bringing
the system to a quantum critical point.

\section{Disorder modeling and technique}
\label{method}

We now turn to the systems considered above with finite concentration $c$ of defects. Here we discuss impurities which change only exchange coupling constants in corresponding Hamiltonians and which do not change the nature of the paramagnetic phase at $H=0$ (i.e., the ground state remains singlet). For instance, we do not consider defects below which weaken intradimer coupling constants so much that local magnetic moments arise on imperfect bonds. On the other hand, we do not assume below that deviation of the coupling constants on imperfect bonds from their values in pure systems is small. Two types of disorder can be distinguished: i) disorder in the intradimer exchange coupling constant $\cal J$ or in the value of the single-ion anisotropy $\cal D$ and ii) disorder in small exchange coupling constants $J_{ij}$ between spins from different dimers or spins on neighboring sites (in large-$\cal D$ systems).

Hamiltonians of systems with defects are written as
\begin{equation}
  \mathcal{H}=\mathcal{H}_2+V,
  \label{impHam}
\end{equation}
where $\mathcal{H}_2$ is given by Eqs.~\eqref{h2}, \eqref{h22}, \eqref{h23}, and \eqref{h24} and $V$ has the following form for disorder in $\cal J$ or $\cal D$ only
\begin{equation}
  V = \sum_{\{n\}} u \mathbf{S}_{n,1} \cdot \mathbf{S}_{n,2},
	\text{ or }
	V = \sum_{\{n\}} u \left(S^z_{n}\right)^2,
	\label{v}
\end{equation}
where the summation is taken over all imperfect bonds or sites and $u$ measures the deviation of $\cal J$ or $\cal D$ on imperfect bonds or sites from their values in pure systems. It is seen from Eqs.~\eqref{spinrep}, \eqref{hp} and \eqref{drep} that such a disorder effects only the chemical potential value on imperfect bonds or sites which is parametrized by the single parameter $u$. Thus, one obtains at $H<H_{c1}$ from Eqs.~\eqref{v} for one sort of particles ($\mathfrak a$-particles, for definiteness)
\begin{equation}
 V = u \sum_{\{n\}}\mathfrak a^+_n \mathfrak a_n.
	\label{v1}
\end{equation}

\begin{figure}
  \noindent
  \hfil
  \includegraphics[scale=0.5]{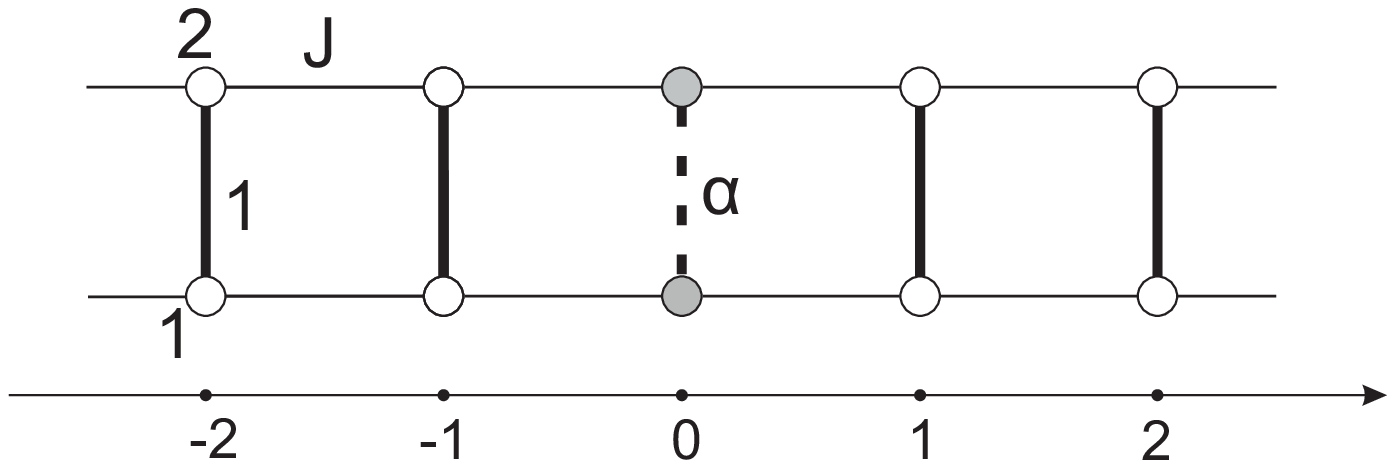}
  \includegraphics[scale=0.5]{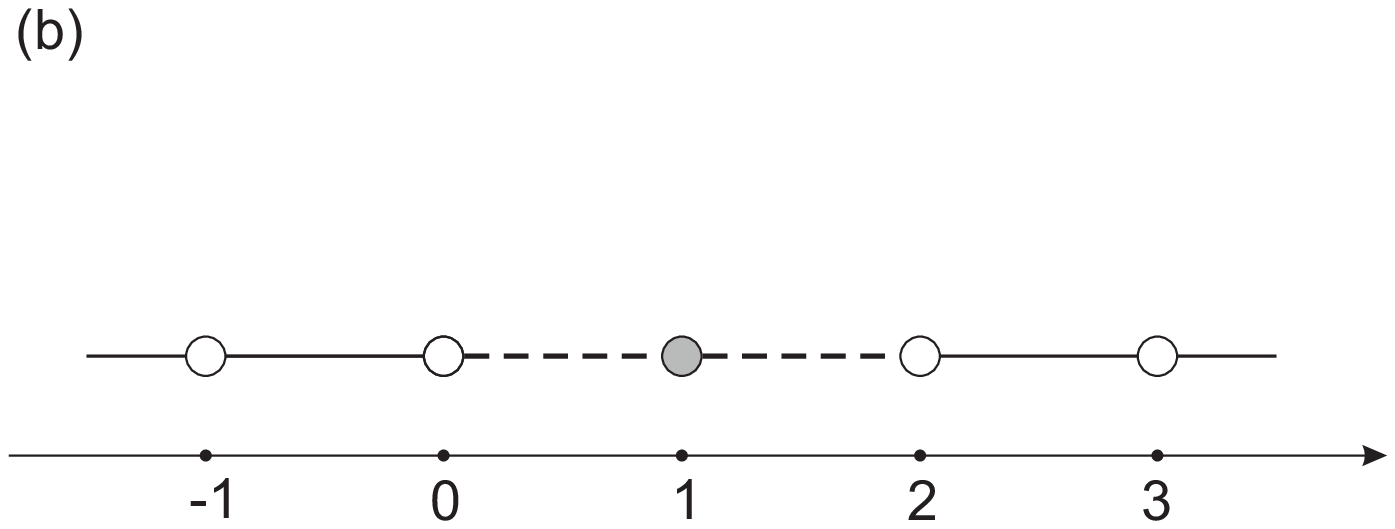}
  \hfil
  \caption{1D systems with imperfect bonds shown by dashed lines. (a) Spin-$\frac12$ ladder with dimers on rungs (shown in bold) with modified intradimer exchange constant $\cal J$ at rung 1 and modified values of exchange coupling constants between spins from dimer 1 and neighboring dimers 0 and 2. (b) Integer spin chain with modified value of the single-ion easy-plain anisotropy $\cal D$ at site 1 and modified value of exchange coupling constant at bonds 0-1 and 1-2.}
  \label{systems1D}
\end{figure}

Expressions for $V$ are cumbersome for disorder in $J_{ij}$ and we present them for 1D systems only which are shown in Fig.~\ref{systems1D}:
\begin{equation}
  \begin{split}
    &V = u_1\sum_{\{n\}} \left(\mathbf{S}_{n,1} \cdot \mathbf{S}_{n+1,1}+ \mathbf{S}_{n,2} \cdot \mathbf{S}_{n+1,2} + \mathbf{S}_{n-1,1} \cdot \mathbf{S}_{n,1} + \mathbf{S}_{n-1,2} \cdot \mathbf{S}_{n,2}\right)\\
	\text{ or }
	  &V = u_1\sum_{\{n\}} \left(\mathbf{S}_{n} \cdot \mathbf{S}_{n+1}+ \mathbf{S}_{n-1} \cdot \mathbf{S}_{n} \right),
  \end{split}
\end{equation}
where the first equation is for the spin-$\frac12$ ladder (see Fig.~\ref{systems1D}(a)), the second one is for the integer spin chain (see Fig.~\ref{systems1D}(b)), $u_1$ measures the deviation of $J_{ij}$ on imperfect bonds from its value in pure systems. The part of the perturbation operators corresponding to $\mathfrak a$-particles has the form
\begin{equation}
  V = \frac {a u_1}{2}\sum_{\{n\}} \left( \mathfrak a^+_n \mathfrak a_{n+1} + \mathfrak a^+_{n+1}\mathfrak a_n + \mathfrak a^+_{n-1} \mathfrak a_n + \mathfrak a^+_n \mathfrak a_{n-1} \right),
	\label{v2}
\end{equation}
where $a=1$ for spin ladder and $a=S(S+1)$ for integer-spin chains if $H=0$. We omit in Eq.~\eqref{v2} terms of the form $\mathfrak b^+_i \mathfrak a^+_j$, $\mathfrak b_i \mathfrak a_j$, $\mathfrak c^+_i \mathfrak c^+_j$, and $\mathfrak c_i \mathfrak c_j$, which arises at small $H$ and give corrections of the next order in $J_{ij}$.

We start our discussion below with the disorder in $\cal J$ or $\cal D$ only. Then, we also add the disorder in $J_{ij}$ and discuss corresponding results for systems with two types of disorder. As it is usually done \cite{Vojta,Zhel13}, we assume that these two types of disorder are "coupled". For example, spins from an imperfect dimer are coupled to spins from neighboring dimers by imperfect bonds (see, e.g., Fig.~\ref{systems1D} for 1D systems). This assumption is quite natural because in real materials substitution of a non-magnetic atom usually changes all exchange coupling constants in its vicinity.

\subsection{$T$-matrix approach}

We use the conventional $T$-matrix approach (see, e.g., Refs.~\cite{Izyumov,white,wan,donia}) to find analytically corrections to quasiparticles spectra and density of states (DOS). Processes involving simultaneous scattering on more than a single impurity are omitted in this technique and all results are valid in the first order in $c$. One obtains for Green's functions of each mode in disordered systems
\begin{equation}
  G(\mathbf{k},E)=\frac{1}{E-\varepsilon_{\mathbf{k}}-cT({\bf k},E)},
  \label{green1}
\end{equation}
where $T({\bf k},E)$ is a quantity related to the $T$-matrix and $\varepsilon_{\mathbf{k}}$ is the pure system quasiparticle spectrum. Then, the translation invariance of systems is effectively restored in the first order in $c$ that allows using Green's functions of the form \eqref{green1} to analyze the spectra of these modes. \cite{Izyumov,white,wan} The quantity $T({\bf k},E)$ can be expressed via coordinate Green's functions of pure systems
\begin{equation}
  G_{nm}(E)=\frac{1}{N} \sum_\mathbf{p} \frac{e^{i \mathbf{p}(\mathbf{R}_n-\mathbf{R}_m) }}{E-\varepsilon_{\mathbf{p}} - i0},
  \label{Green1}
\end{equation}
where $N$ is the number of unit cells.

For disorder in $\cal J$ or $\cal D$ only, $T({\bf k},E)$ does not depend on momentum having the form
\begin{equation}
  T({\bf k},E)=\frac{u}{1- uG_{00}(E)},
  \label{Wk1}
\end{equation}
where $u$ measures the deviation of $\cal J$ or $\cal D$ on imperfect bonds or sites from their values in pure systems (see Eq.~\eqref{v}). Spectrum of quasiparticles $E_\mathbf{k}$ and their damping $\gamma_\mathbf{k}$ are defined by poles of Green's function \eqref{green1} and they have the form in the first order in $c$
\begin{equation}
 E_\mathbf{k}=\varepsilon_{\mathbf{k}}+c \Re{(T({\bf k},E=\varepsilon_{\mathbf{k}}))},
\quad
\gamma_\mathbf{k}=c \Im{(T({\bf k},E=\varepsilon_{\mathbf{k}}))},
  \label{encor}
\end{equation}
where $\Re$ and $\Im$ denote real and imaginary parts, respectively. Eqs.~\eqref{encor} are written under assumption that the solution of the equation $E-\varepsilon_{\mathbf{k}}-cT({\bf k},E)=0$ at fixed $\bf k$ can be expanded as series in $c$ in which the first terms taken into account in Eqs.~\eqref{encor} are much larger than higher order terms. It can happen, however, that this is not the case for some $\bf k$. It would signify that diagrams of higher orders in $c$ have to be taken into account and Eqs.~\eqref{green1}--\eqref{encor} have to be reconsidered. As we obtain below, it is the situation that arises in considered systems for states in excitation band lying near its bottom and the top, where the following inequalities should hold for Eqs.~\eqref{encor} validity:
\begin{eqnarray}
\label{valid}
	&&|\varepsilon_{\mathbf{k}}-\Delta| \gg c|T({\bf k},\varepsilon_{\mathbf{k}})|,\\
\label{valid2}
	&&|\varepsilon_{\mathbf{k}}-\Delta-a|J_{\bf k_0}|| \gg c|T({\bf k},\varepsilon_{\mathbf{k}})|,
\end{eqnarray}
respectively. To analyze states near bands edges, we perform numerical calculations discussed below in details. It should be noted that invalidity of Eqs.~\eqref{green1}--\eqref{encor} and the necessity to go beyond the first order in $c$ is usually seen from an analysis similar to that just described. It happens sometimes that processes of multiple-defects scattering are important and their contributions (which are of higher orders in $c$) are much larger than the first order corrections. We demonstrate below that it is the situation which arises in 1D systems under discussion.

Defects modify the system DOS \cite{Izyumov,white} $g(E)$. The general expression for $g(E)$ has the following form in the first order in $c$ in the case of disorder in $\cal J$ or $\cal D$ only:
\begin{equation}
  g(E) = g_0(E) - c\frac{u^2 g_0(E) \Re (dG_{00}/dE)+u(1-u \Re (G_{00}(E)))dg_0/dE}{\left[ (1-u \Re (G_{00}(E)))^2 + (\pi u g_0(E))^2\right]},
  \label{ds1}
\end{equation}
where $g_0(E)=\Im(G_{00}(E))/\pi$ is the pure system DOS and $G_{00}(E)$ is given by Eq.~\eqref{Green1} with $m=n=0$. It is seen from Eq.~\eqref{ds1} that the correction to DOS can have extrema when the following condition is satisfied:
\begin{equation}
  1-u \Re(G_{00}(E))=0.
  \label{ds2}
\end{equation}
It is well-known that in magnetically ordered phases solutions of equations similar to Eq.~\eqref{ds2} give positions of isolated levels (localized states) outside the excitation band (where $g_0(E)=0$) or virtual levels (resonances) inside the band. \cite{Izyumov,white} However, we find below that Eq.~\eqref{ds2} gives only positions of isolated impurity levels in the paramagnetic phases and all anomalies inside the band stem from derivatives in the numerator of the second term in Eq.~\eqref{ds1}.

Imperfection in $J_{ij}$ can be taken into account in the same way although the corresponding analytical consideration is more technically involved. Some details on this point can be found in Appendix~\ref{append} devoted to 1D systems. Green's functions of propagating modes have the form \eqref{green1}, where $T({\bf k},E)$ does depend on momentum $\bf k$ and Green's functions \eqref{Green1} with $m\ne n$ also contribute to it. As a result expressions for $T({\bf k},E)$ and DOS are more cumbersome than Eqs.~\eqref{Wk1} and \eqref{ds1} and we do not present them here although the spectrum renormalization is given in the first order in $c$ by Eqs.~\eqref{encor} as before.

\subsection{Numerical calculations}
\label{numcalc}

To confirm our analytical results and to reveal the nature of states near excitation bands edges, we perform numerical diagonalization of the one-particle sector of the bosonic Hamiltonians \eqref{impHam}, \eqref{v1} and \eqref{v2} for finite 1D and 2D systems with disorder. Thus, we find eigenvalues, eigenfunctions and DOS $\rho(\epsilon)$ of finite systems. Energy and damping of a propagating mode with momentum $\bf k$ are found using the Green's function definition
\begin{equation}
G({\bf k},t) = -i\langle vac |T\mathfrak a_{\bf k}(t)\mathfrak a^\dagger_{\bf k}(0) |vac\rangle
= -i\langle {\bf k} | e^{-i{\cal H}t} |{\bf k}\rangle \theta(t)
= -i\sum_\epsilon \left|\langle {\bf k} | \epsilon \rangle\right|^2e^{-i\epsilon t} \theta(t),
\label{g}
\end{equation}
where $|vac\rangle$ is the ground state of the Hamiltonian $\cal H$ given by Eqs.~\eqref{impHam}, \eqref{v1} and \eqref{v2}, $|{\bf k}\rangle$ is the state with a particle having momentum $\bf k$ (plane wave), $\theta(t)$ is the Heaviside step function, and $|\epsilon \rangle$ is the eigenfunction of $\cal H$ corresponding to eigenvalue $\epsilon$. One can replace the summation on $\epsilon$ by integration inside the band and we have from Eq.~\eqref{g}
\begin{equation}
G({\bf k},\omega) = \int d\epsilon\frac{f({\bf k},\epsilon)}{\omega-\epsilon+i0}
= \dashint d\epsilon\frac{f({\bf k},\epsilon)}{\omega-\epsilon}-i\pi f({\bf k},\omega),
\label{g22}
\end{equation}
where $f({\bf k},\epsilon)=\rho(\epsilon)|\langle {\bf k} | \epsilon \rangle|^2$ can be found from the exact diagonalization results and $\dashint$ denotes the principal value of the integral. It follows from Eq.~\eqref{g22} that $f({\bf k},\omega)$ is related to the imaginary part of the Green's function $G({\bf k},\omega)$ which should have the Lorentzian shape for a well-defined propagating quasiparticle with momentum $\bf k$. Thus, one can obtain the energy and the damping of propagating excitations by fitting $f({\bf k},\omega)$ with the Lorentzian.

To characterize quantitatively the spatial localization/delocalization of a state $\psi$ found by diagonalization, we calculate also the inverse participation ration (IPR):
\begin{equation}
\label{iprexpr}
\mbox{IPR}(\psi) = \sum_n |\psi(n) |^4,
\end{equation}
where $n$ labels the lattice sites. IPR is of the order of the inverse number of sites occupied at state $\psi$. Then, IPR scales as $1/L^d$ for spatially extended states and it is equal to a constant for localized states, where $d$ is the system dimension. Exponential localization is characterized by ${\rm IPR}\propto 1/\xi^d$, where $\xi$ is of the order of localization length (see, e.g., Ref.~\cite{Vojta}).

The number of sites in considered clusters vary from 400 to 15000. For each cluster, we perform an averaging over a large number of disorder realizations to find $f({\bf k},\epsilon)$, DOS and IPR. The number of disorder realizations vary from $10^5$ for the smallest clusters to 600 for the largest ones. We try both periodic and open boundary conditions which lead to the same results. Corrections to the quasiparticles energy and their damping are found by an extrapolation of numerical data for a number of finite size systems containing $L^d$ unit cells to thermodynamic limit using quadratic polynomials in $1/L$. In particular, Fig.~\ref{1dspec} presented in the next section for 1D systems is build using 250 momentum values. Extrapolations in planes \ref{1dspec}(a) and \ref{1dspec}(b) are carried out using $L=1024$, 2048, 4096, and 8192. $f({\bf k},\epsilon)$ shown in insets are calculated for $L=6144$. In planes \ref{1dspec}(c) and \ref{1dspec}(d), clusters with $L=500$, 1000, 2000, and 3000 are used for extrapolations and insets show results for $L=3000$.

\section{Disordered systems}
\label{disordersec}

\subsection{1D systems}

\subsubsection{$T$-matrix approach}

\begin{figure}
  \noindent
	\includegraphics[scale=0.4]{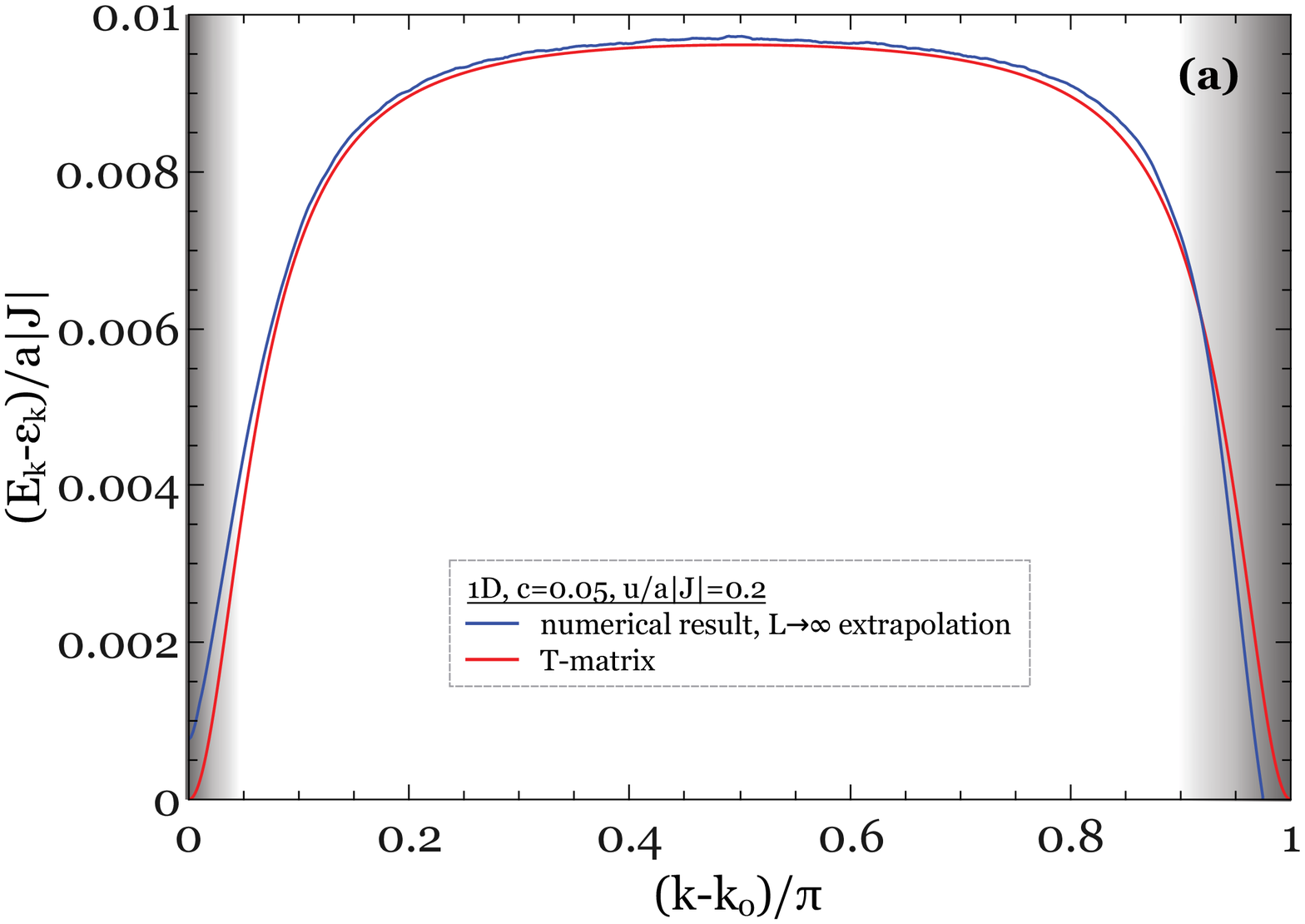}
	\includegraphics[scale=0.4]{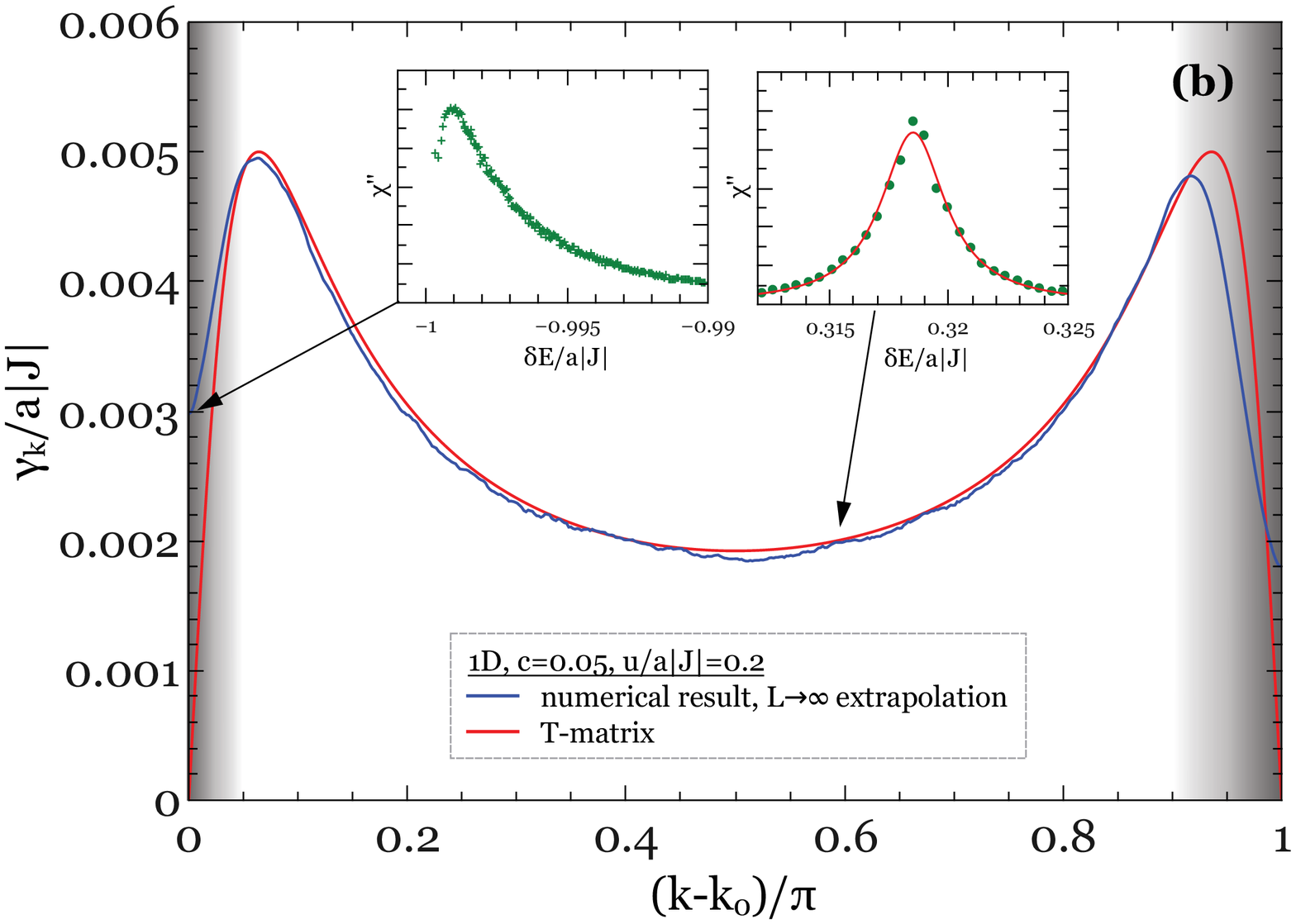}
	\includegraphics[scale=0.4]{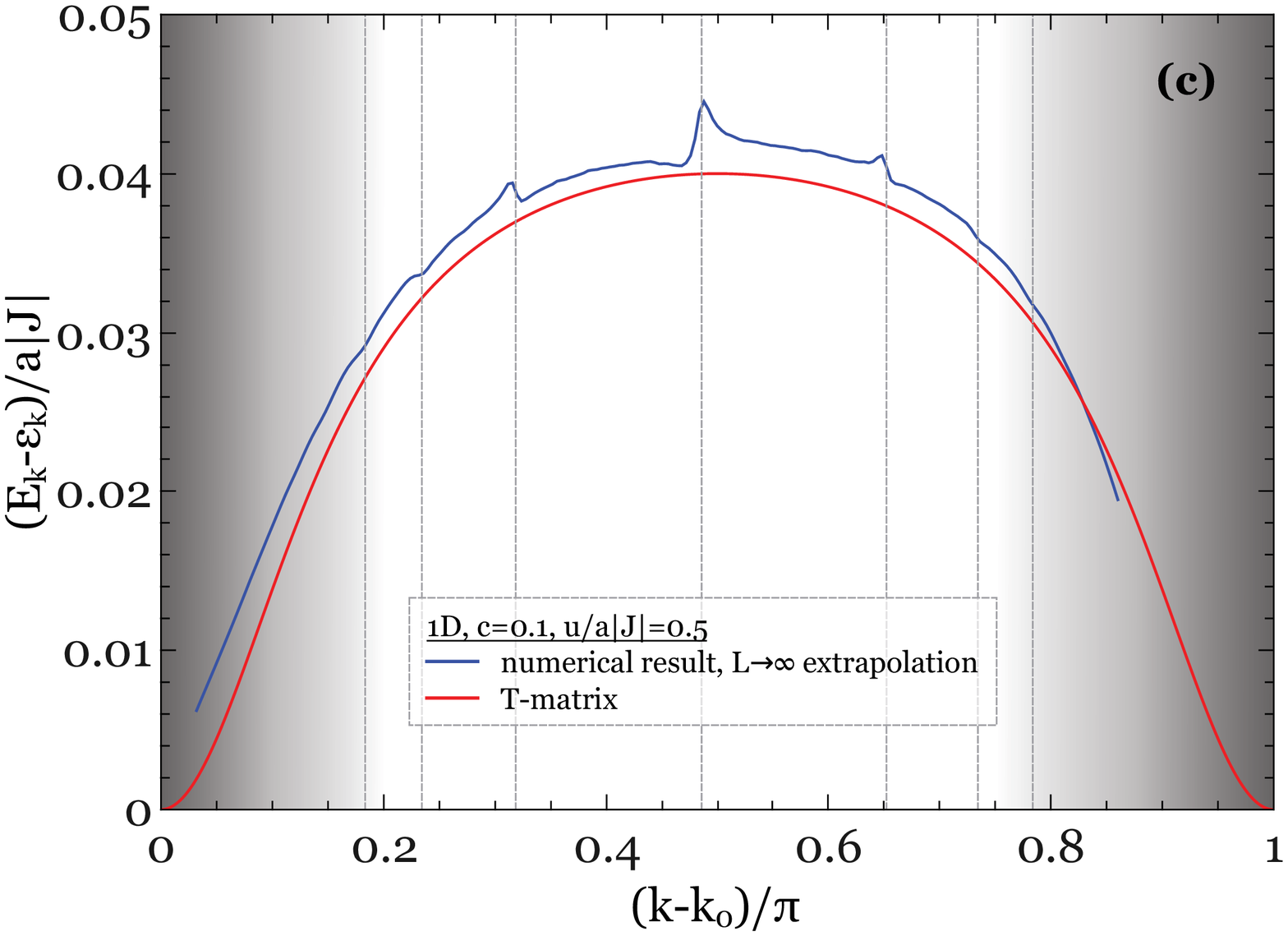}
  \includegraphics[scale=0.4]{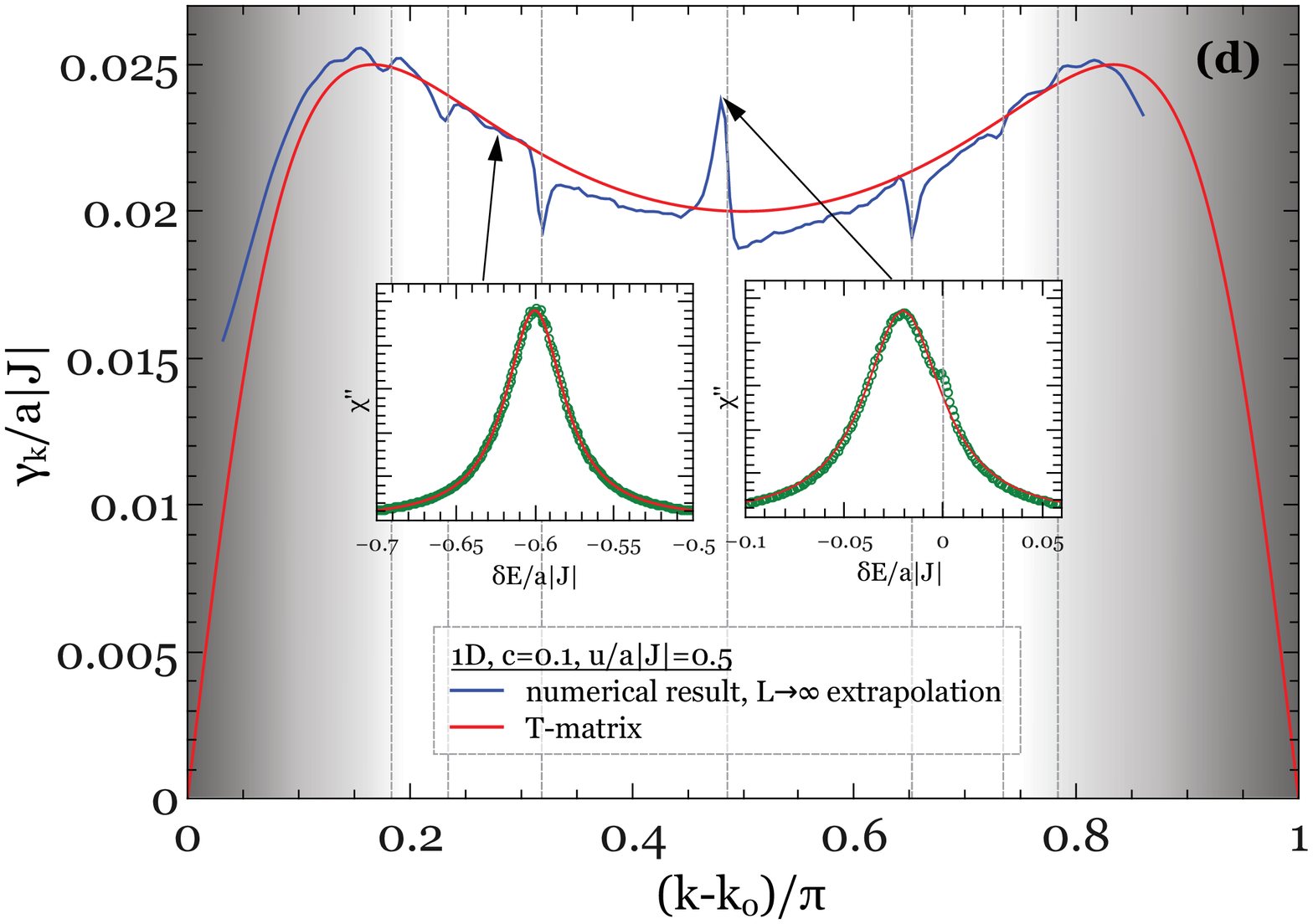}
  \hfil
  \caption{
	(Color online.) Correction to quasiparticles energy and their damping in 1D systems given by Eqs.~\eqref{spec1d} and found numerically for disorder in $\cal J$ or $\cal D$ only (the extrapolation is carried out of numerical data for finite systems containing $L$ unit cells to thermodynamical limit as it is explained in Sec.~\ref{numcalc}). Shaded regions mark areas in which the imaginary parts of one-particle Green's functions $\chi''(k,\omega)$ found numerically using Eq.~\eqref{g22} do not have a Lorentzian shape and in which our analytical results are invalid (i.e., inequalities \eqref{val1d} do not hold). Insets in planes (b) and (d) show $\chi''(k,\omega)$ for some fixed momenta, where solid lines represent results of data fitting by Lorentzians. Most pronounced anomalies in numerical data for ``stronger'' impurities (planes (c) and (d)) are interpreted as a result of coherent scattering by defects of quasiparticles with momenta denoted by vertical lines (see the text).
	}
  \label{1dspec}
\end{figure}

Calculations are particularly simple in 1D systems with defects which are depicted in Fig.~\ref{systems1D}. Taking into account the exchange coupling between nearest neighbors only, one has for the bare spectrum (cf. Eq.~\eqref{spec})
\begin{equation}
  \varepsilon_{\bf k} = \Delta+a|J|+aJ\cos k,
	\label{1dspec2}
\end{equation}
where $\Delta = 1-a|J|$ at $H=0$ and the spectrum minimum is located at $k=k_0$, where $k_0=\pi$ and 0 for $J>0$ and $J<0$, respectively. We obtain after simple integration in Eq.~\eqref{Green1}
\begin{equation}
  G_{00}(E)= \begin{cases}
    \frac{1}{\sqrt{(E-\Delta-a|J|)^2-a^2J^2}}, & E>\Delta+2a|J|, \\
    \frac{i}{\sqrt{a^2J^2-(E-\Delta-a|J|)^2}}, & \Delta< E\le \Delta+2a|J|, \\
    -\frac{1}{\sqrt{(E-\Delta-a|J|)^2-a^2J^2}}, & E<\Delta.
		\label{g001d}
  \end{cases}
\end{equation}
Using the second line in Eq.~\eqref{g001d}, one has
$
  G_{00}(E=\varepsilon_{k})=i/|aJ \sin{k}|
$
and we obtain for the spectrum and the damping from Eqs.~\eqref{Wk1} and \eqref{encor} in the case of disorder in $\cal J$ or $D$ only
\begin{equation}
  E_k = \Delta+a|J|+aJ\cos k + c \frac{ua^2J^2 \sin^2 k}{a^2J^2 \sin^2 k+u^2},
\qquad
  \gamma_k=c \frac{u^2 a|J \sin{k}|}{a^2J^2 \sin^2 k+ u^2}.
\label{spec1d}
\end{equation}
Let us discuss the neighborhood of the spectrum minimum, where it has the form \eqref{spec0}. It is seen from Eqs.~\eqref{spec1d} that there are two regimes at $\kappa=|k-k_0|\ll1$:
\begin{equation}
\begin{array}{llll}
  E_k &= \Delta + \left(\frac{a|J|}{2} + c\frac{a^2J^2}{u} \right)\kappa^2,
	\qquad
	&\gamma_k = ca|J|\kappa,
	\quad
	&\mbox{if}\quad  \kappa\ll \min\{1,|u/aJ|\},\\
  E_k &= \Delta + cu + \frac{a|J|}{2} \kappa^2,
	\qquad
	&\gamma_k = c\frac{u^2}{a|J|\kappa},
	\quad
	&\mbox{if}\quad  1\gg\kappa\gg |u/aJ|.
  \label{E1d}
	\end{array}
\end{equation}
The range of Eqs.~\eqref{E1d} validity given by Eq.~\eqref{valid} reads
\begin{equation}
\begin{array}{ll}
	\kappa\gg c,\qquad &\text{if}\quad |u|\gg ca|J|,\\
		\kappa\gg \sqrt{c\left|\frac{u}{aJ}\right|}, \qquad &\text{if}\quad |u|\ll ca|J|.
\end{array}
	\label{val1d}
\end{equation}
One is lead from Eq.~\eqref{valid2} to the same range of the analytical approach validity near the spectrum maximum (in this case $\kappa$ measures a deviation of momentum from the value at which the spectrum has the maximum). Correction to the quasiparticle energy and its damping given by Eqs.~\eqref{spec1d} are plotted in Fig.~\ref{1dspec} for particular parameters values. The corresponding numerical results are also shown in Fig.~\ref{1dspec} which are discussed below.

\begin{figure}
  \noindent
  \includegraphics[scale=0.4]{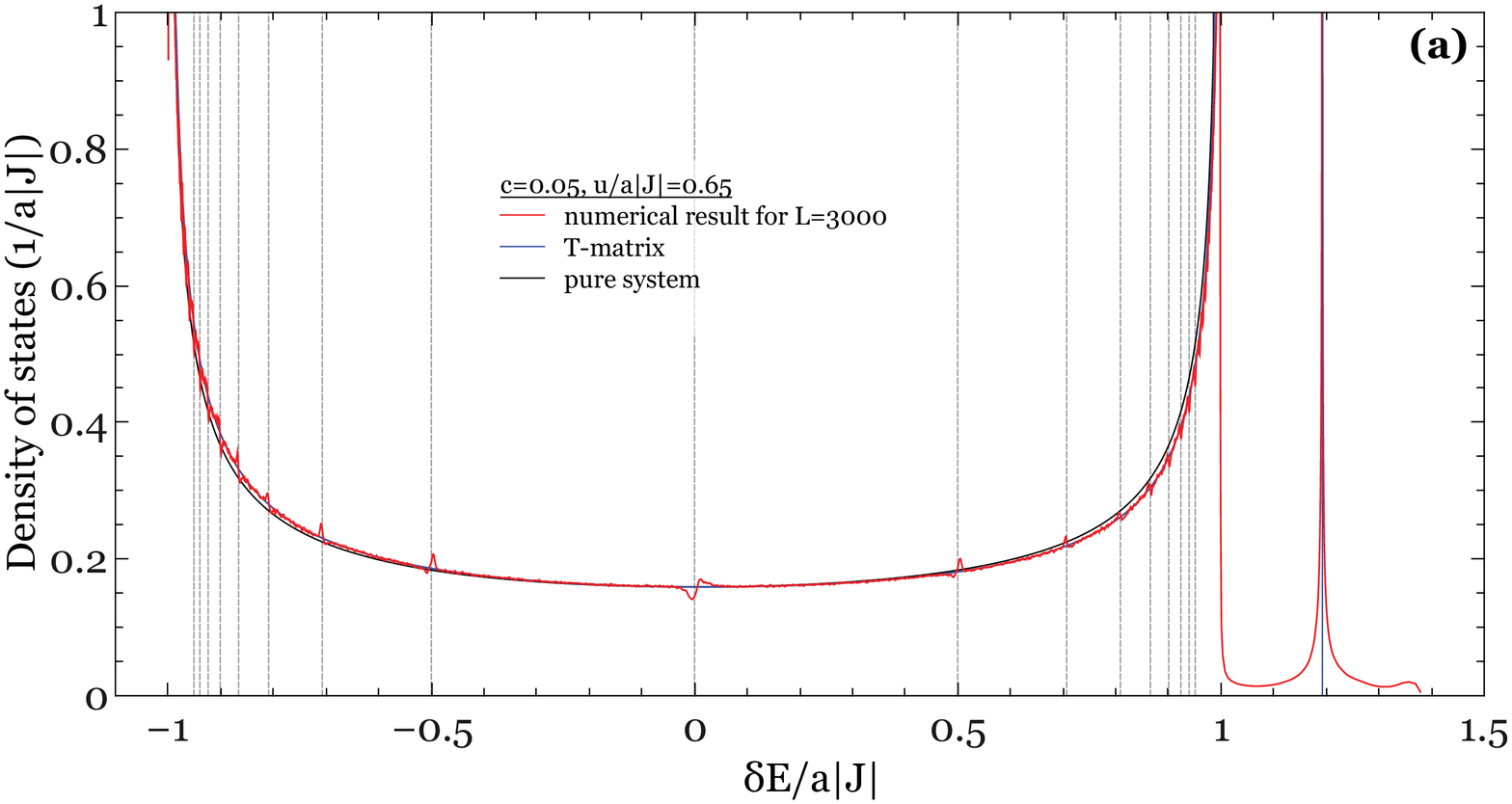}
  \includegraphics[scale=0.4]{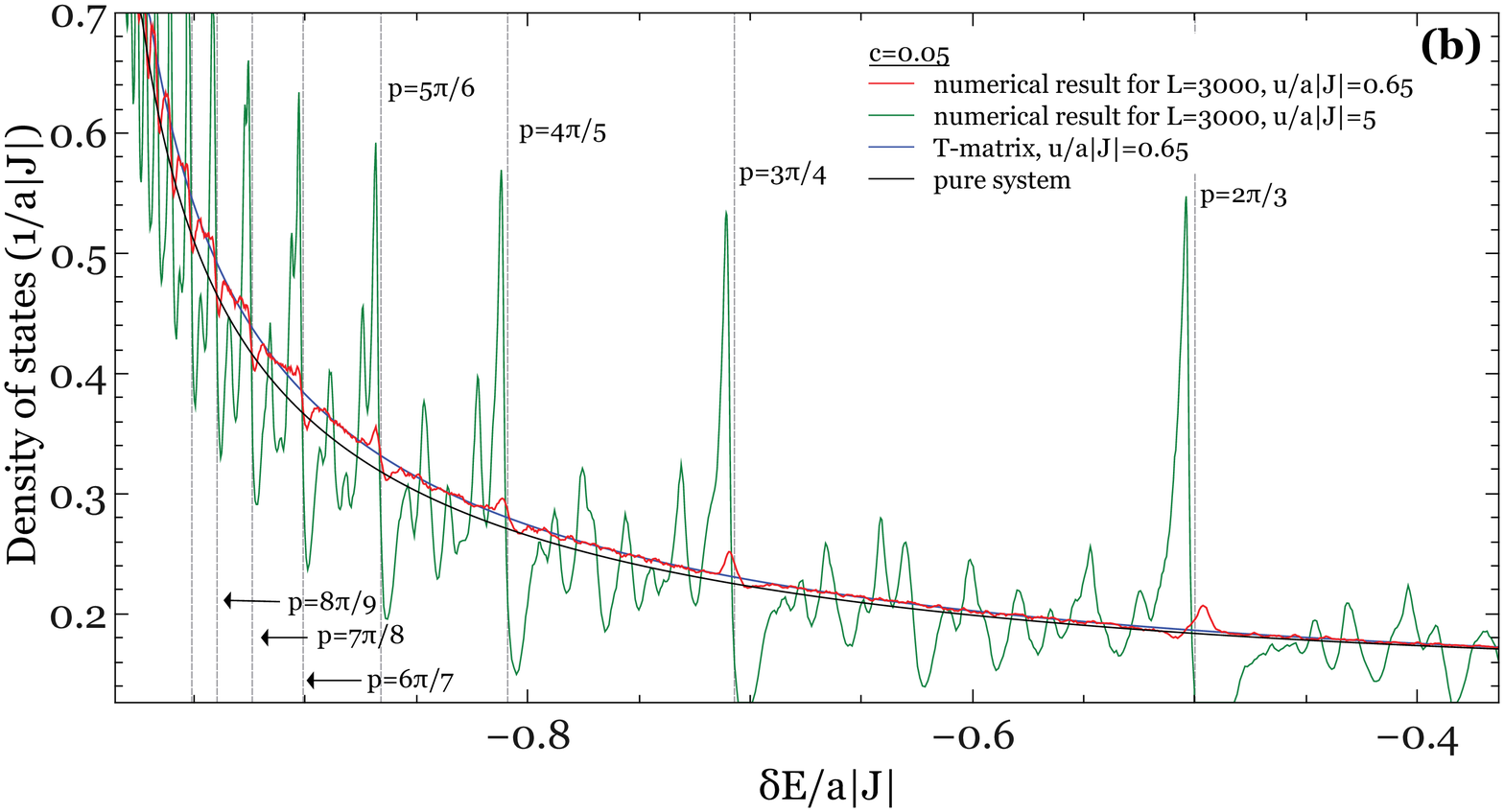}
  \hfil
  \caption{(Color online.) DOS of 1D systems with disorder in $\cal J$ or $\cal D$ only, where $\delta E=E-\Delta-a|J|$. DOS in the $T$-matrix approach is given by Eq.~\eqref{ds1} (it is almost indistinguishable on the plots from the pure system DOS). Most pronounced anomalies in numerical data found for $L=3000$ are interpreted as a result of coherent scattering by defects of quasiparticles with energies denoted by vertical lines (see the text).}
  \label{dos1}
\end{figure}

DOS of pure systems $g_0(E)$ is equal to $G_{00}(E)/i\pi=\frac1\pi(a^2J^2-(E-\Delta-a|J|)^2)^{-1/2}$ inside the band. One concludes from Eq.~\eqref{ds1} that defects do not lead to noticeable corrections to DOS in the range of the analytical approach validity determined by Eqs.~\eqref{val1d}. Outside the band, where $G_{00}(E)$ is real, an isolated impurity level appears above or below the band depending on the sign of $u$. One has from Eqs.~\eqref{ds1}, \eqref{ds2}, and \eqref{g001d} for $E$ lying outside the band (see Fig.~\ref{dos1}(a))
\begin{equation}
  g(E) = c \delta(E-E_d), \qquad E_d = \Delta+a|J|+{\rm sign}(u) \sqrt{a^2J^2+u^2}.
	\label{imp1d}
\end{equation}
Multiple-impurities scattering processes, which are not taken into account in the first order in $c$, turn this isolated level into a narrow impurity band. Eqs.~\eqref{imp1d} are in accordance with the corresponding result of Ref.~\cite{Arlego} devoted mainly to DOS in disordered spin-$\frac12$ ladders.

Imperfection in the small exchange coupling $J_{ij}$ is considered in detail in Appendix~\ref{append}. Eq.~\eqref{w1d2} is derived there for $T(k,E)$ that gives for the spectrum using Eqs.~\eqref{encor} (cf.\ Eqs.~\eqref{spec1d})
\begin{eqnarray}
  E_k &=& \Delta+a|J|+aJ\cos k + c\frac{\left( 1+\frac{u_1}{J}\right)^2\left(u+2 a u_1\cos k + a u^2_1\cos k/J\right)a^2 J^2 \sin^2 k}{\left( 1+ \frac{u_1}{J}\right)^4a^2 J^2 \sin^2 k + \left(u+2 a u_1\cos k +a u^2_1\cos k/J\right)^2},\nonumber\\
	\gamma_k &=& ca|J \sin k| \frac{\left(u+2 a u_1\cos k + a u^2_1\cos k/J\right)^2}{\left( 1+ \frac{u_1}{J}\right)^4 a^2 J^2 \sin^2 k + \left(u+2 a u_1\cos k +a u^2_1\cos k/J\right)^2}.
	\label{spec1d22}
\end{eqnarray}
If $|u|J|-2 a u_1 J- au^2_1|\gg a(J+u_1)^2 |\sin k|$, the spectrum has the form near its minimum (cf.\ Eqs.~\eqref{E1d})
\begin{equation}
  E_k=\Delta+\left(\frac{a|J|}{2} + c\frac{a^2J^2(1+u_1/J)^2}{u-2 a u_1 J/|J|- au^2_1/|J|} \right)\kappa^2,
	\qquad
	\gamma_k=ca|J|\kappa.
	\label{spec1d2}
\end{equation}
We point out also the reduction of the spectrum renormalization by two sorts of disorder when $u|J|-2 a u_1 J- au^2_1\approx0$ and $|1+u_1/J|\sim1$. One obtains in this case from Eqs.~\eqref{spec1d22} $|E_k-\varepsilon_k|\sim c|J|\kappa^2\ll\varepsilon_k-\Delta$ and $\gamma_k\sim c|J|\kappa^3\ll\varepsilon_k-\Delta$.

DOS in 1D systems with two sorts of disorder is also considered in Appendix~\ref{append}. It is shown there, in particular, that there are no isolated impurity levels at $a|u_1|\gg|u|$ if $-2<u_1/J<0$, whereas one level above and one level below the band arise if $u_1/J$ lies outside this interval (in accordance with Ref.~\cite{Arlego}).

It should be noted that Eqs.~\eqref{g001d}, \eqref{spec1d}, and \eqref{spec1d22} are derived using the particular form of the spectrum \eqref{1dspec2}. One would lead to different results for gapped phases with another spectrum. In contrast, Eqs.~\eqref{E1d}--\eqref{val1d} and \eqref{spec1d2} are more universal because they can be obtained using the general form of the spectrum \eqref{spec0} near its minimum (or maximum) and the form of the impurity interaction \eqref{v1} and \eqref{v2} (the combination $a|J|$ in these expressions stems from the factor in the expression \eqref{spec0} for the spectrum and $au_1$ originates from $V$ given by Eq.~\eqref{v2}). This is due to the fact that small $\kap_p$ give the main contribution to $G_{nm}$ in Eq.~\eqref{Green1}, where $\kap_p$ is the deviation of $\bf p$ from the momentum at which the spectrum has minimum (or maximum).

\subsubsection{Numerical results}
\label{num1d}

Our numerical results for the quasiparticle energy, damping and DOS are also presented in Figs.~\ref{1dspec} and \ref{dos1} (for the disorder in $\cal J$ or $\cal D$ only). As it is seen, they are in good agreement with analytical findings in the range of the analytical approach validity \eqref{val1d} except for some points near which upward and downward spikes appear. Amplitudes of these spikes rise as $|u|$ and/or $c$ increase. As the $T$-matrix approach does not show such anomalies, we attribute them to resonances in multiple scattering on defects which are not taken into account in our analytical consideration and which are effects of higher order in $c$. The origin of these resonances can be understood qualitatively by noting that elementary excitations of a pure chain with momenta $k=m\pi/n$ and $k=\pi-m\pi/n$, where $m<n$ are integers, scatter coherently by defects which are $rn$ sites apart, where $r$ is integer. If the renormalized spectrum $E_k$ differs noticeably from the bare spectrum $\varepsilon_k$ given by Eq.~\eqref{1dspec2}, positions of anomalies shift a little due to the fact that an excitation with energy $E_k$ produces excitations with the same energy $\varepsilon_p=E_k$ as a result of scattering on defects which interfere coherently if $p=m\pi/n$. Positions of resonances found in this way are denoted in Figs.~\ref{1dspec}(c), \ref{1dspec}(d) and \ref{dos1} by vertical lines which mark accurately location of anomalies in numerical data (momenta $p$ are also depicted in Fig.~\ref{dos1}(b) near corresponding vertical lines).

The imaginary part of the one-particle Green's function $\chi''(k,\omega)$ is shown for some momenta in insets of Figs.~\ref{1dspec}(b) and \ref{1dspec}(d) which have been found numerically as it is described above. These insets illustrate our finding that $\chi''(k,\omega)$ has the Lorentzian shape for not too strong impurities in the range of the analytical approach validity. Upon $u$ and/or $c$ increasing, amplitudes of anomalies rise and the form of peaks in $\chi''(k,\omega)$ in the vicinity of corresponding $k$ bears little resemblance to a Lorentzian for large enough $u$. The resonant scattering becomes strong enough and our analytical results are completely invalid when $c|u/aJ|\agt1$ (see Fig.~\ref{dos1}(b)).


Peaks in $\chi''(k,\omega)$ have non-Lorentzian shapes near the band edges for all $u$ and $c\ll1$ (see Fig.~\ref{1dspec}(b) for illustration). Areas in $k$-space with non-Lorentzian peaks are shaded in Fig.~\ref{1dspec} that illustrates our results for $|u/aJ|\sim1$ when anomalies inside the band are not too large. These areas widths are in accordance with our estimations \eqref{val1d} of regions sizes in which analytical results are valid.

\begin{figure}
  \noindent
  \includegraphics[scale=0.4]{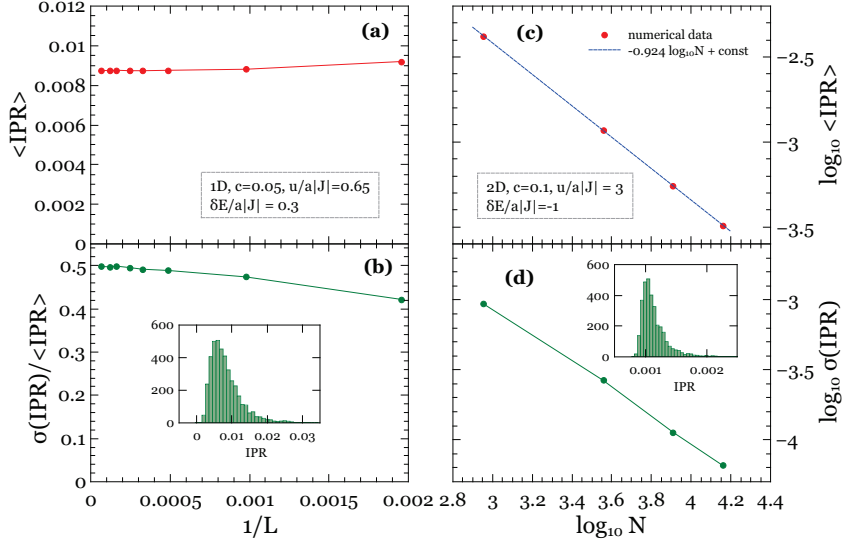}
  \hfil
  \caption{(Color online.) The inverse participation ration (IPR) given by Eq.~\eqref{iprexpr} averaged over disorder realizations as it is described in Sec.~\ref{numcalc}. $\sigma(\rm IPR)$ is the mean square deviation of IPR from its mean value. (a), (b) and (c), (d) slides are for particular states inside the band far from its edges in 1D and 2D systems, respectively. The states energies $\delta E$ are measured from the band center (see Figs.~\ref{dos1}(a) and \ref{dos2}, correspondingly). Insets show histograms of IPR distributions in disorder realizations.
	}
  \label{iprfig}
\end{figure}

Our analysis of IPR defined by Eq.~\eqref{iprexpr} demonstrates that all the states in the band are localized for any $u$ and $c$. This is illustrated by Fig.~\ref{iprfig}(a) drawn for one state inside the band far from its edges. Data are averaged over disorder realizations and the mean square deviation of IPR from its mean value is shown in Fig.~\ref{iprfig}(b). Interestingly, states inside the band far from its edges can combine the localization and properties of short-wavelength wavepacket (if the resonant scattering is not too strong). This situation holds even in the limit $u\to+\infty$ and $u_1\to-J$, when defects break a system to pieces of mean length $1/c$. As is seen from Eqs.~\eqref{spec1d}--\eqref{val1d} and \eqref{spec1d22}, $c$-corrections vanish in expressions for short-wavelength quasiparticles energies but a finite damping remains that reflects a finite lifetime of wavepackets excited in such a broken system.

\subsection{2D systems}

\begin{figure}
  \noindent
  \includegraphics[scale=0.7]{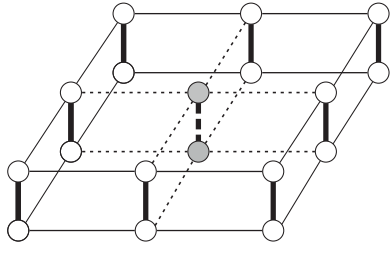}
  \hfil
  \caption{Spin-1/2 dimerized bilayer with imperfect bonds. Notations are the same as in Fig.~\ref{systems1D}(a).}
  \label{SB1}
\end{figure}

We turn to 2D systems with the exchange coupling between nearest neighbors (see Fig.~\ref{SB1} for 2D dimer system) which spectrum has the form
\begin{equation}
  \varepsilon_{\bf k} = \Delta + 2a|J| + aJ(\cos k_x + \cos k_y),
\end{equation}
where $\Delta = 1-2a|J|$ at $H=0$. One obtains taking integral in Eq.~\eqref{Green1} for energies $E>\Delta+4a|J|$ lying outside the band
\begin{equation}
  G_{00}(E)=\frac{2}{\pi(E-\Delta-2a|J|)} K\left(\frac{4a^2J^2}{(E-\Delta-2a|J|)^2} \right),
  \label{Green2D}
\end{equation}
where $K(k)=\int^{\pi/2}_0 \frac{d \theta}{\sqrt{1-k^2 \sin^2 \theta}}$ is the complete elliptic integral of the first kind.

For energies inside the band, $E>\Delta+2a|J|$, the result can be represented in the form
\begin{equation}
  G_{00}(E)=\frac{1}{\pi a|J|} \left[ \frac{1}{\cos \psi} F\left(\frac{\pi}{2}-\psi, \frac{1}{\cos \psi} \right) + \frac{i}{\sin \psi} F\left(\psi, \frac{1}{\sin \psi} \right) \right],
  \label{G02D}
\end{equation}
where $\psi=\arccos \left( \frac{E-\Delta-2a|J|}{2a|J|}\right)$, $F(\phi, k)=\int^{\phi}_0 \frac{d \theta}{\sqrt{1-k^2 \sin^2 \theta}}$ is incomplete elliptic integral of the first kind, and both of the elliptic functions are real.
For other $E$ values, $G_{00}(E)$ can be easily found from Eqs.~\eqref{Green2D} and \eqref{G02D} by using the fact that its real and imaginary parts are antisymmetric and symmetric functions with respect to the point $E=\Delta+2a|J|$, correspondingly. Eq.~\eqref{G02D} can be simplified considerably at $E=\varepsilon_{\mathbf{k}}$ near the spectrum minimum ($\kappa=|k-k_0|\ll1$):
\begin{equation}
   G_{00}(\varepsilon_{\mathbf{k}})\approx \frac{1}{\pi a|J|} \ln\frac{\kappa}{b_{21}} + \frac{i}{2a|J|},
	\label{g002}
\end{equation}
where $b_{21}=\exp{(C_{20})}=2^{5/2}$, $C_{20}$ is a model dependent coefficient,
\begin{equation}
  C_{20}=\pi a |J| \lim_{{\bf k}_1\rightarrow {\bf k}_0} \left( \frac{1}{(2\pi)^2} \int_{\Omega} \frac{d^2k}{\varepsilon_{\bf k} - \varepsilon_{{\bf k}_1}} + \ln k_1\right).
  \label{C_20}
\end{equation}
Using Eqs.~\eqref{Wk1}, \eqref{encor} and \eqref{g002} we have in the vicinity of the spectrum minimum for the disorder in $\cal J$ or $\cal D$ only
\begin{equation}
  E_{\bf k} = \Delta+\frac{a|J|}{2}\kappa^2+c \frac{\pi a|J|(\pi a|J|-u\ln(\kappa/b_{21}))u}{(\pi a|J|-u\ln(\kappa/b_{21}))^2+(\pi u/2)^2},
	\quad
	\gamma_{\bf k} = c \frac{\pi^2 }{2}\frac{a|J|u^2}{(\pi a|J|-u\ln(\kappa/b_{21}))^2+(\pi u/2)^2}.
  \label{2dcorbot}
\end{equation}
One obtains in the same way for the spectrum near the top of the band (i.e., near the spectrum maximum)
\begin{equation}
  E_{\bf k} = \Delta+4a|J|-\frac{a|J|}{2}\kappa^2+c \frac{u\pi a|J|(\pi a|J|+u\ln(\kappa/b_{21}))}{(\pi a|J|+u\ln(\kappa/b_{21}))^2+(\pi u/2)^2},
	\quad
	\gamma_{\bf k} = c \frac{\pi^2 }{2}\frac{u^2a|J|}{(\pi a|J|+u\ln(\kappa/b_{21}))^2+(\pi u/2)^2}.
	 \label{2dcortop}
\end{equation}
The range of Eqs.~\eqref{2dcorbot} and \eqref{2dcortop} validity is written as
\begin{equation}
\begin{array}{cc}
	\kappa\gg\sqrt c, &\mbox{if  } |u|\gg a|J|,\\
	\kappa\gg\sqrt{c\left|\frac{u}{aJ}\right|}, &\mbox{if  } |u|\ll a|J|,
\end{array}
\label{val2d}
\end{equation}
where $\kappa$ measures a deviation of momentum from the values at which the spectrum has maximum or minimum. Notice that all corrections to the spectrum depend weakly on momenta in the range of the results validity: $|E_{\bf k}-\varepsilon_{\bf k}|\sim c$ and $\gamma_{\bf k}\sim c$.

\begin{figure}
  \noindent
  \hfil
  \includegraphics[scale=0.3]{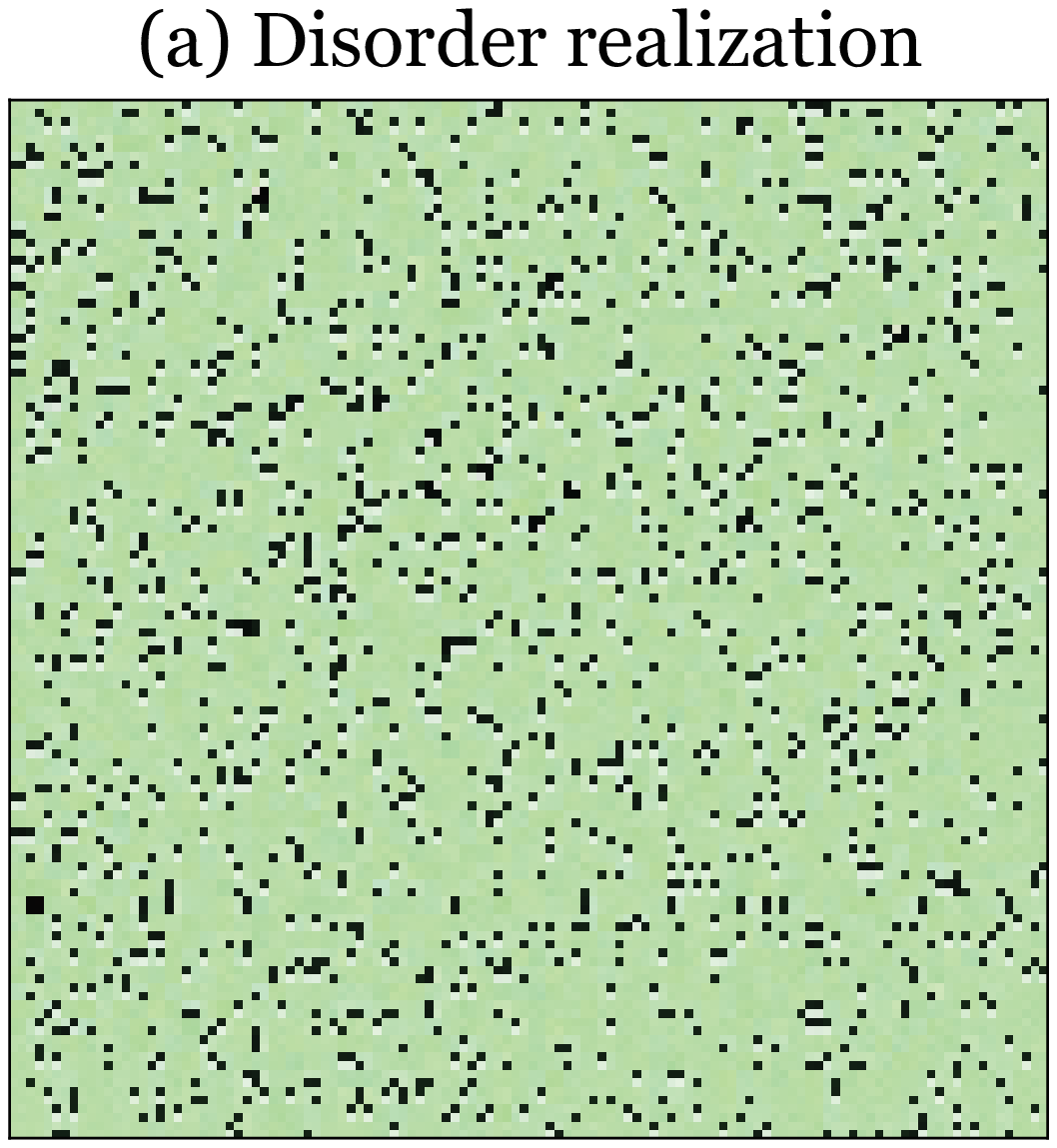}
  \includegraphics[scale=0.3]{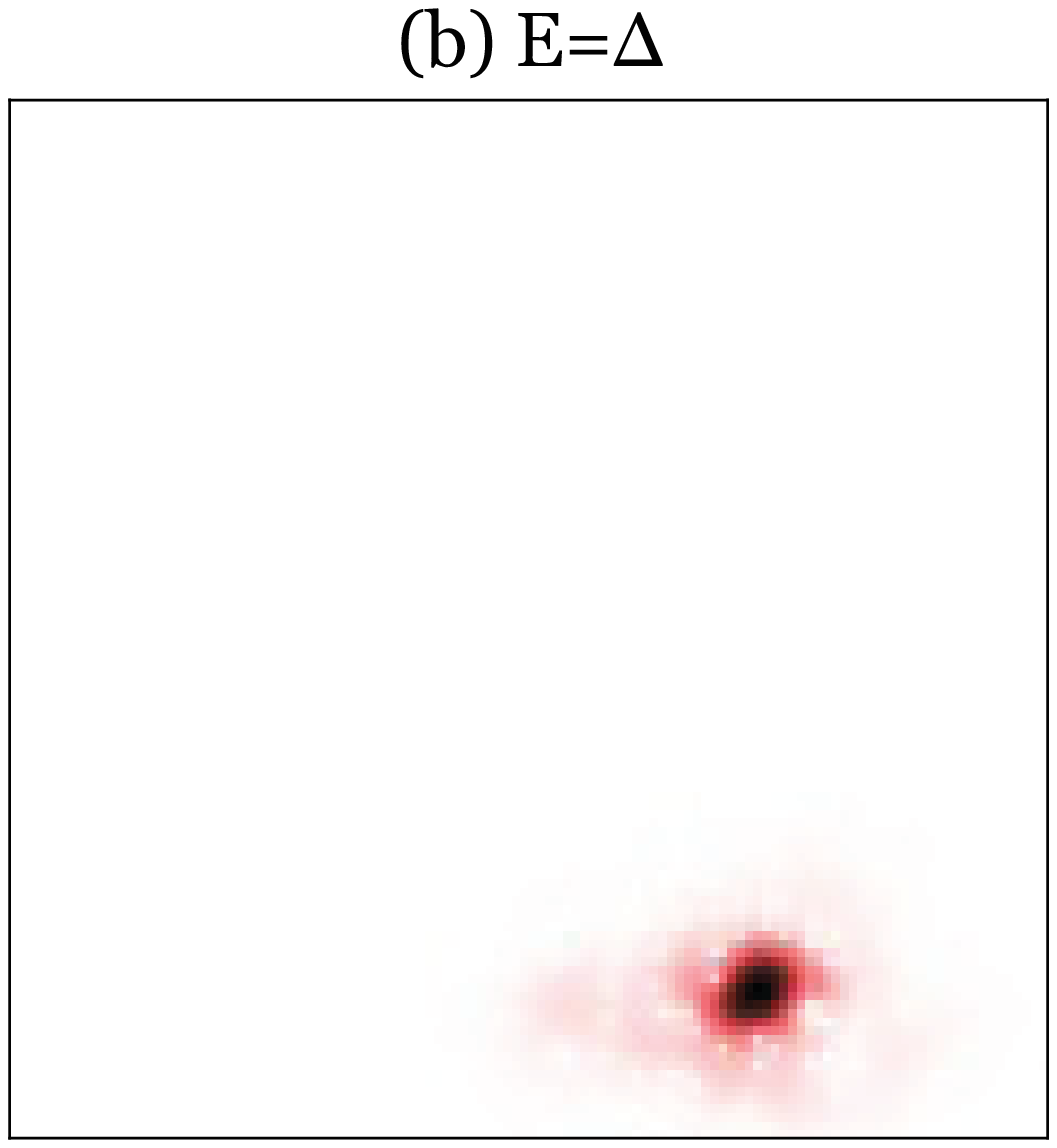}
	\includegraphics[scale=0.3]{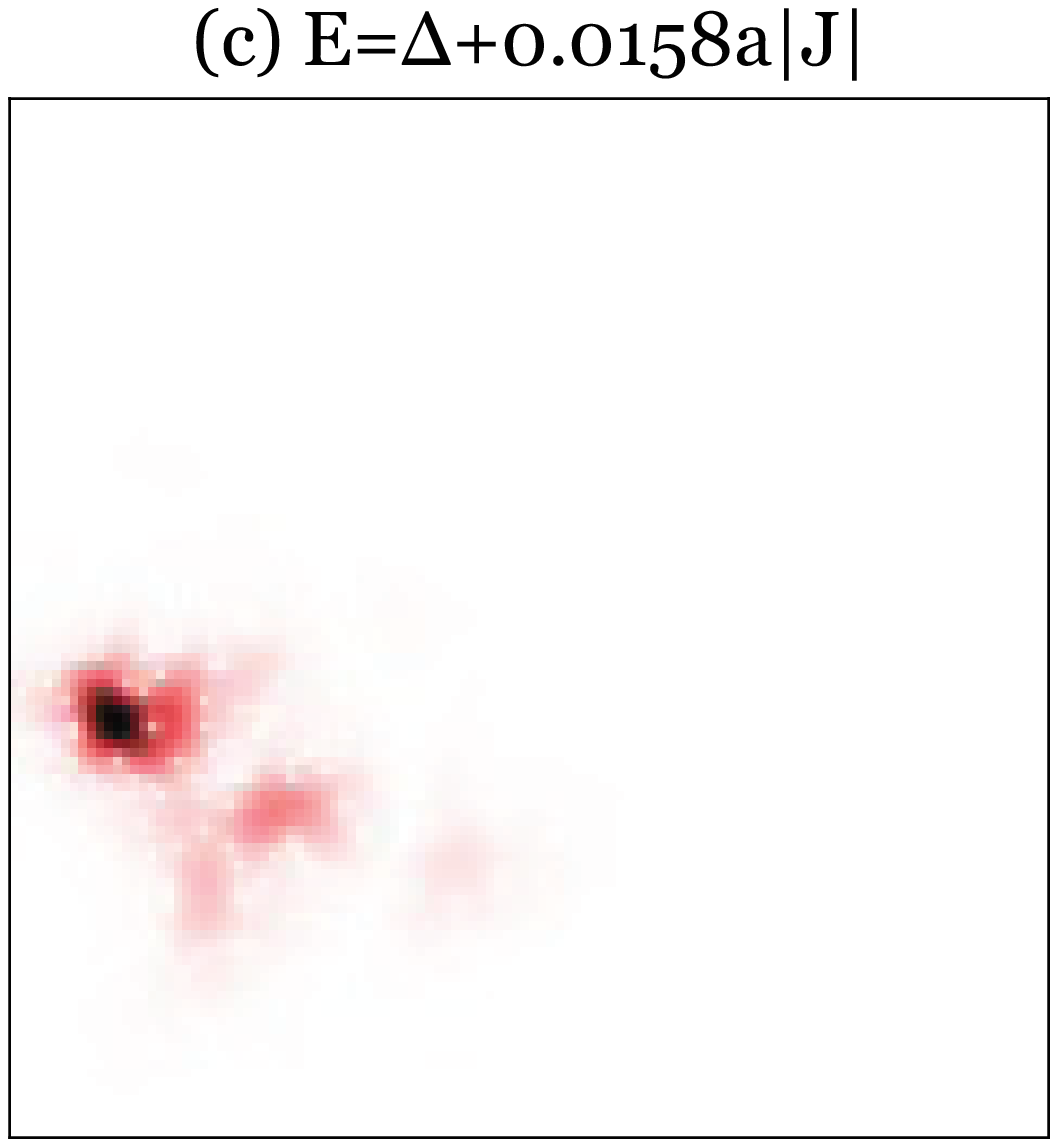}
	\includegraphics[scale=0.3]{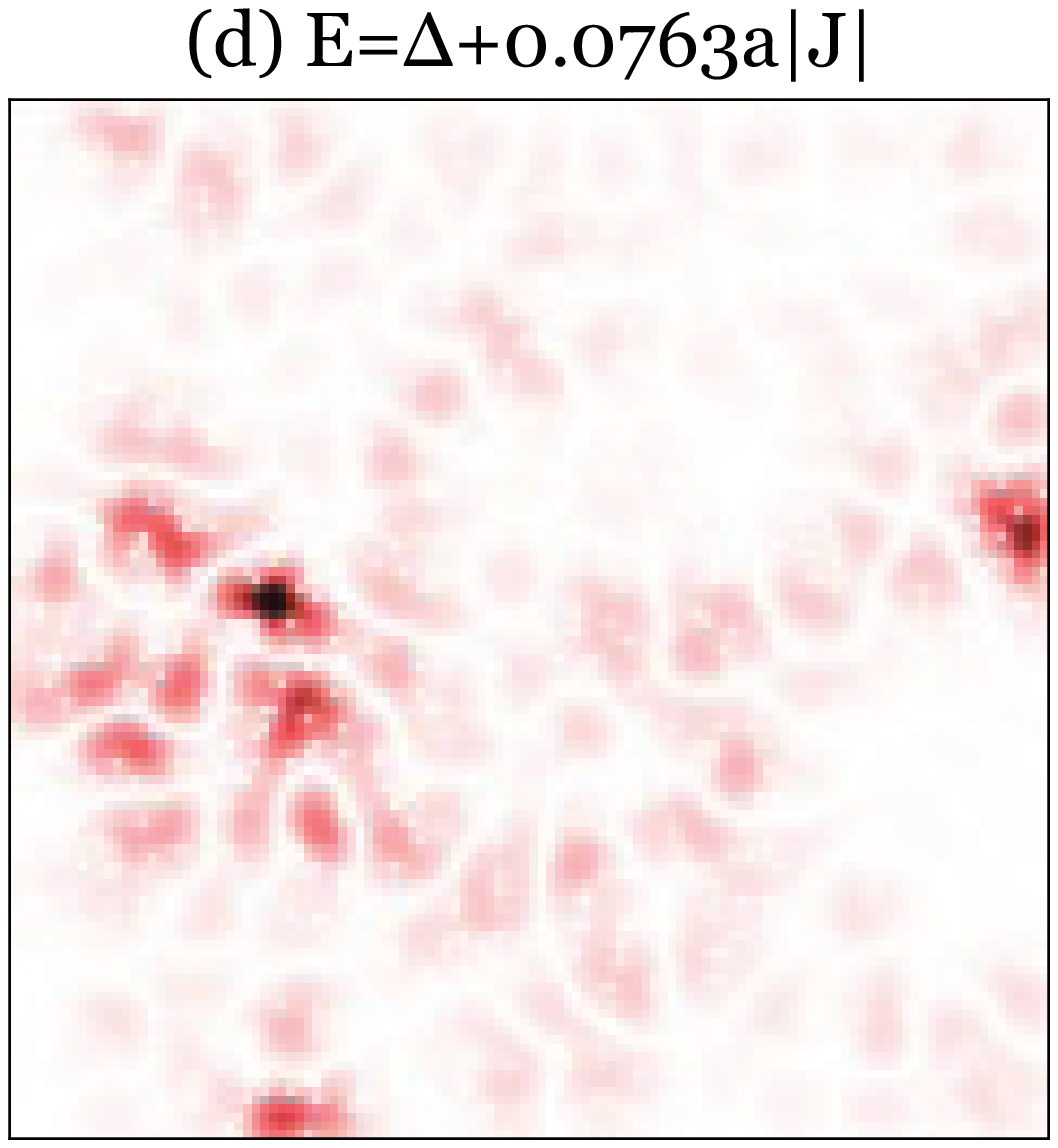}
	\includegraphics[scale=0.3]{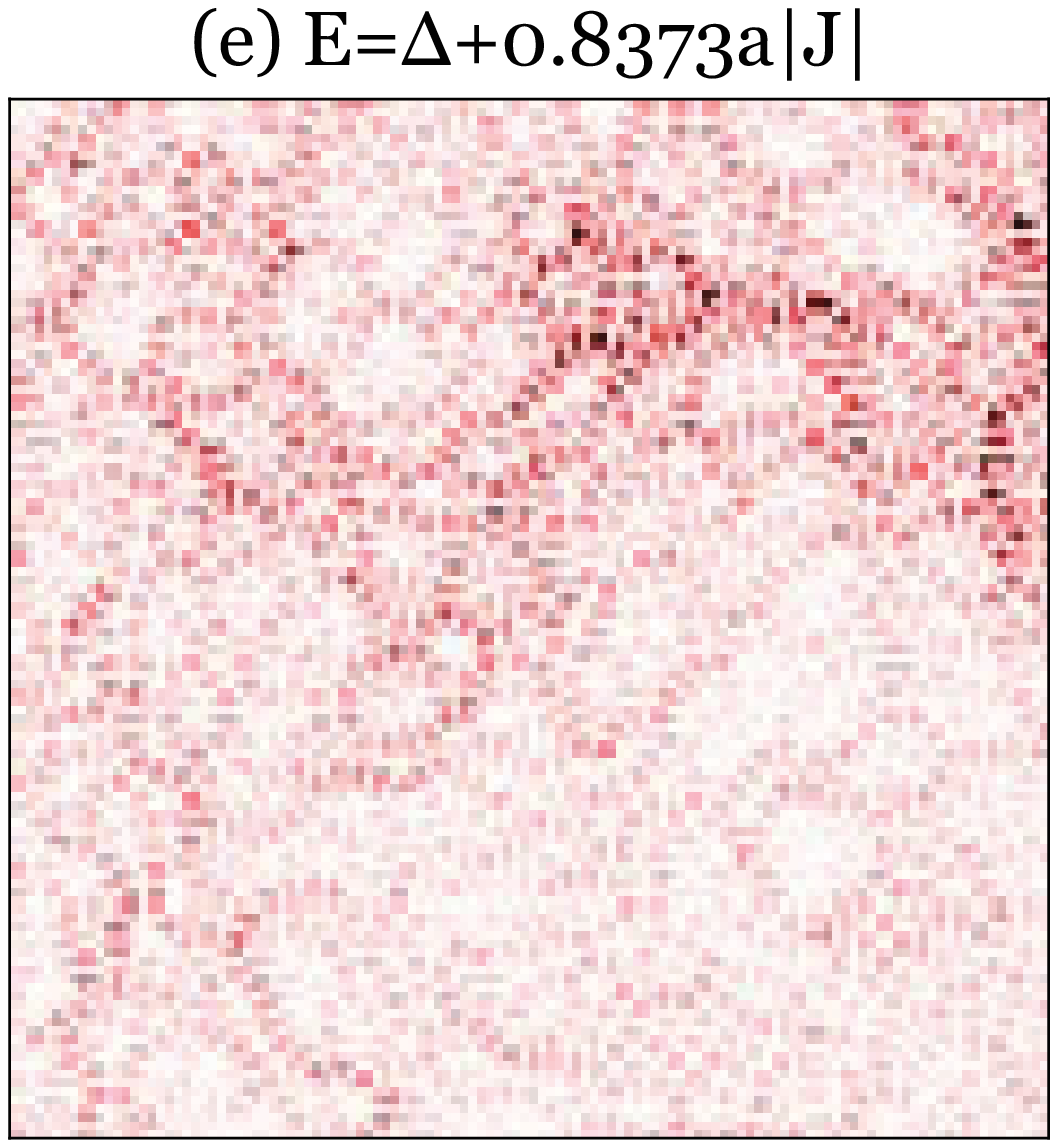}
	\includegraphics[scale=0.3]{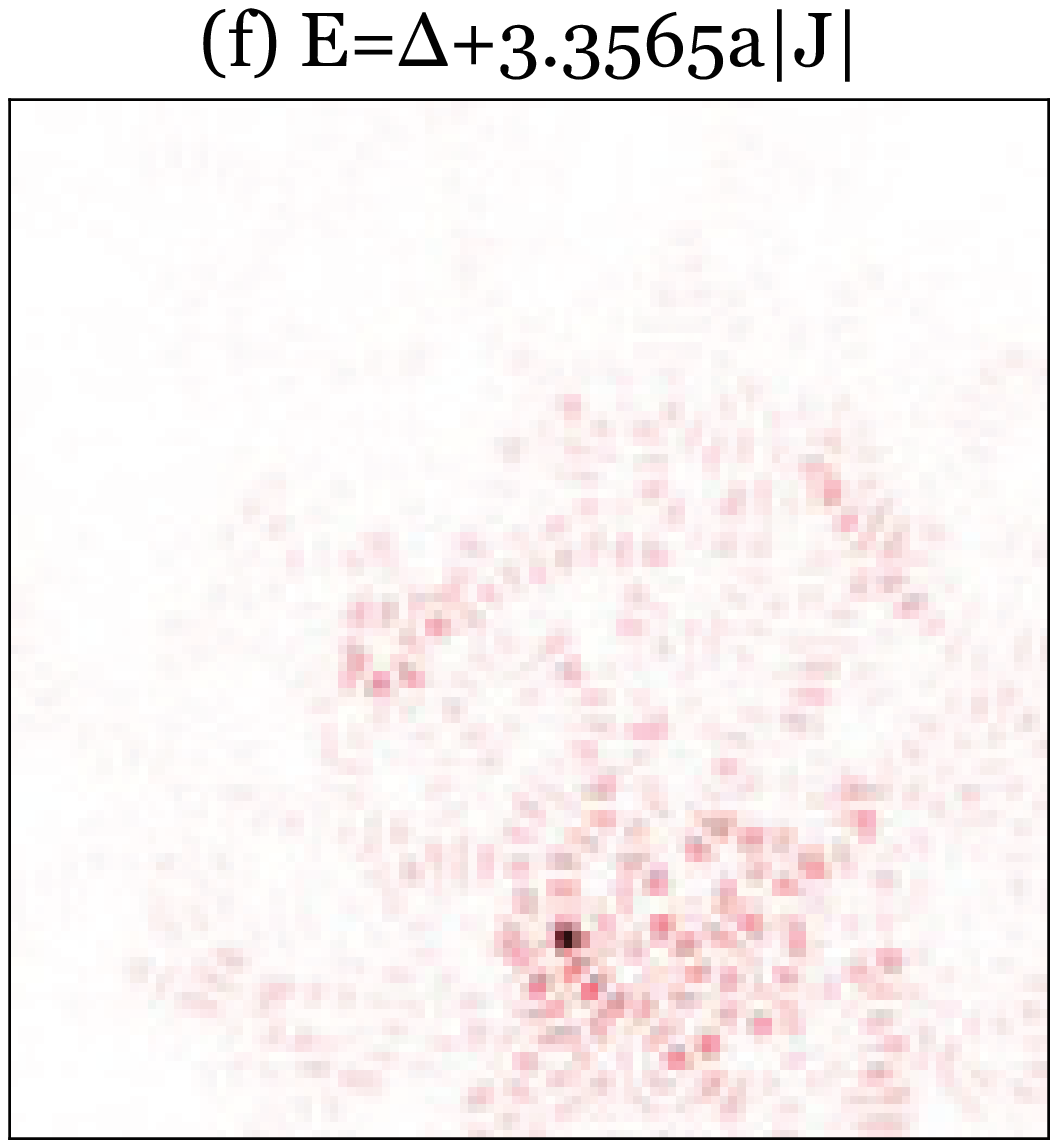}
	\includegraphics[scale=0.3]{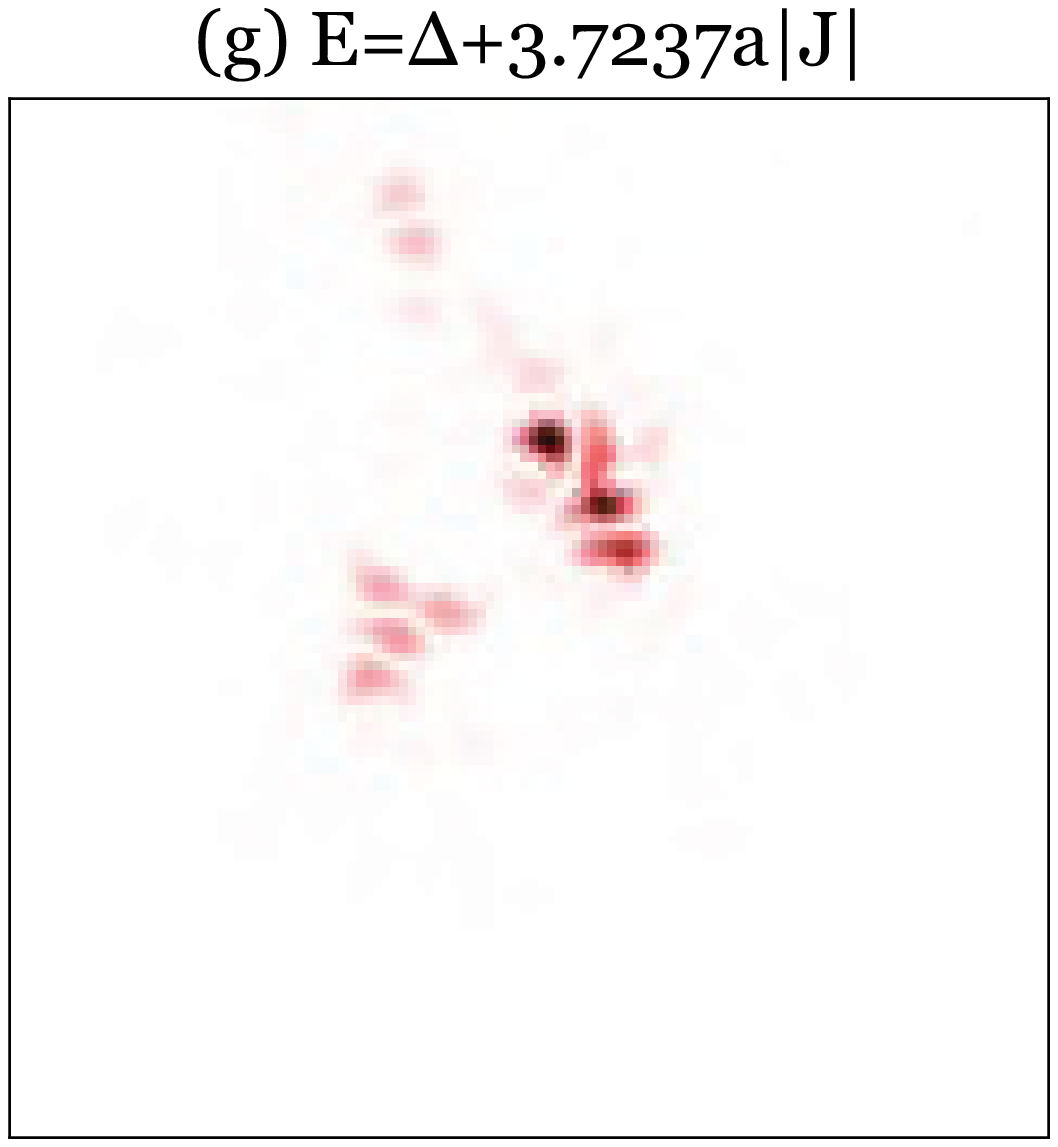}
	\includegraphics[scale=0.3]{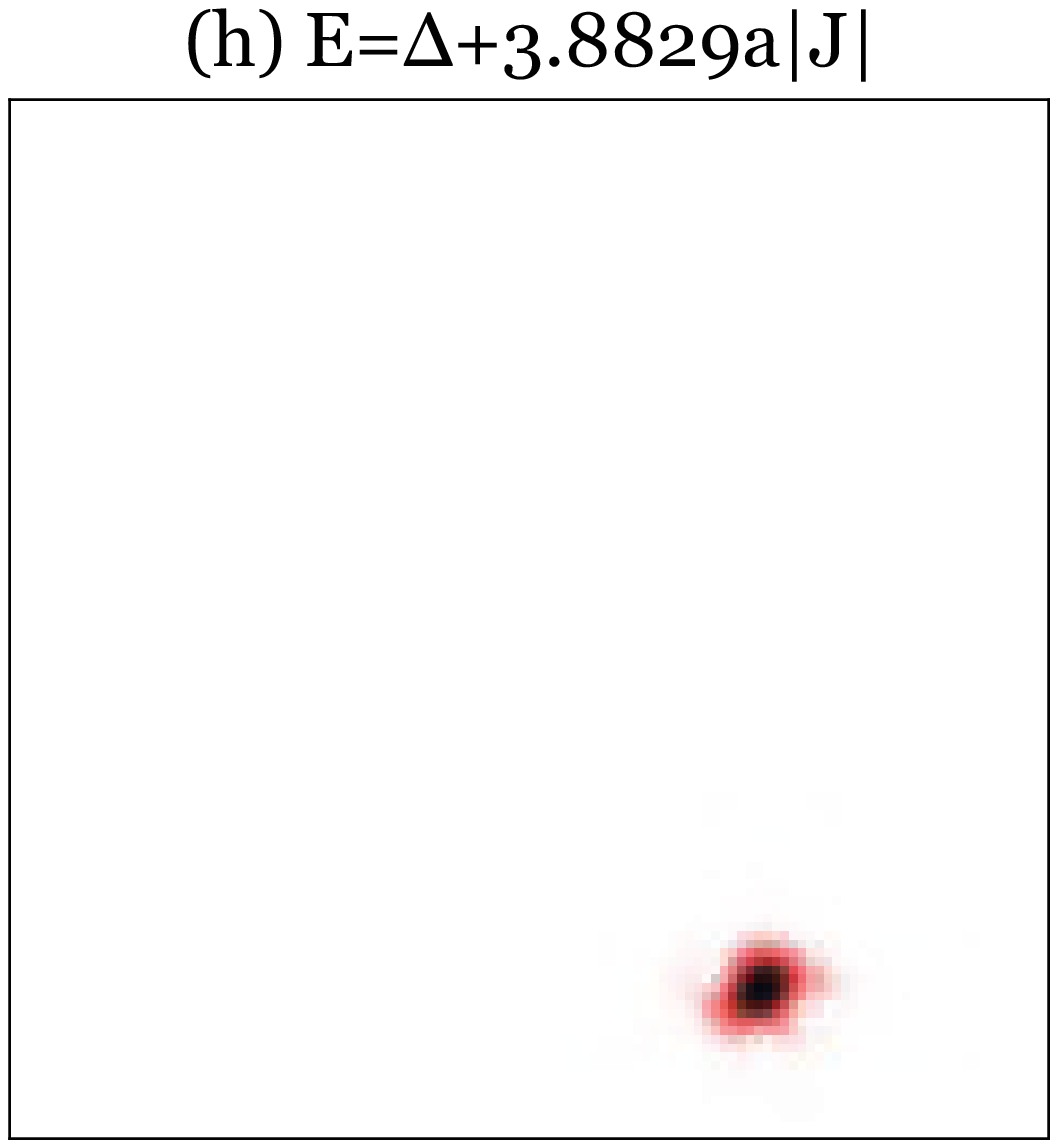}
	\includegraphics[scale=0.3]{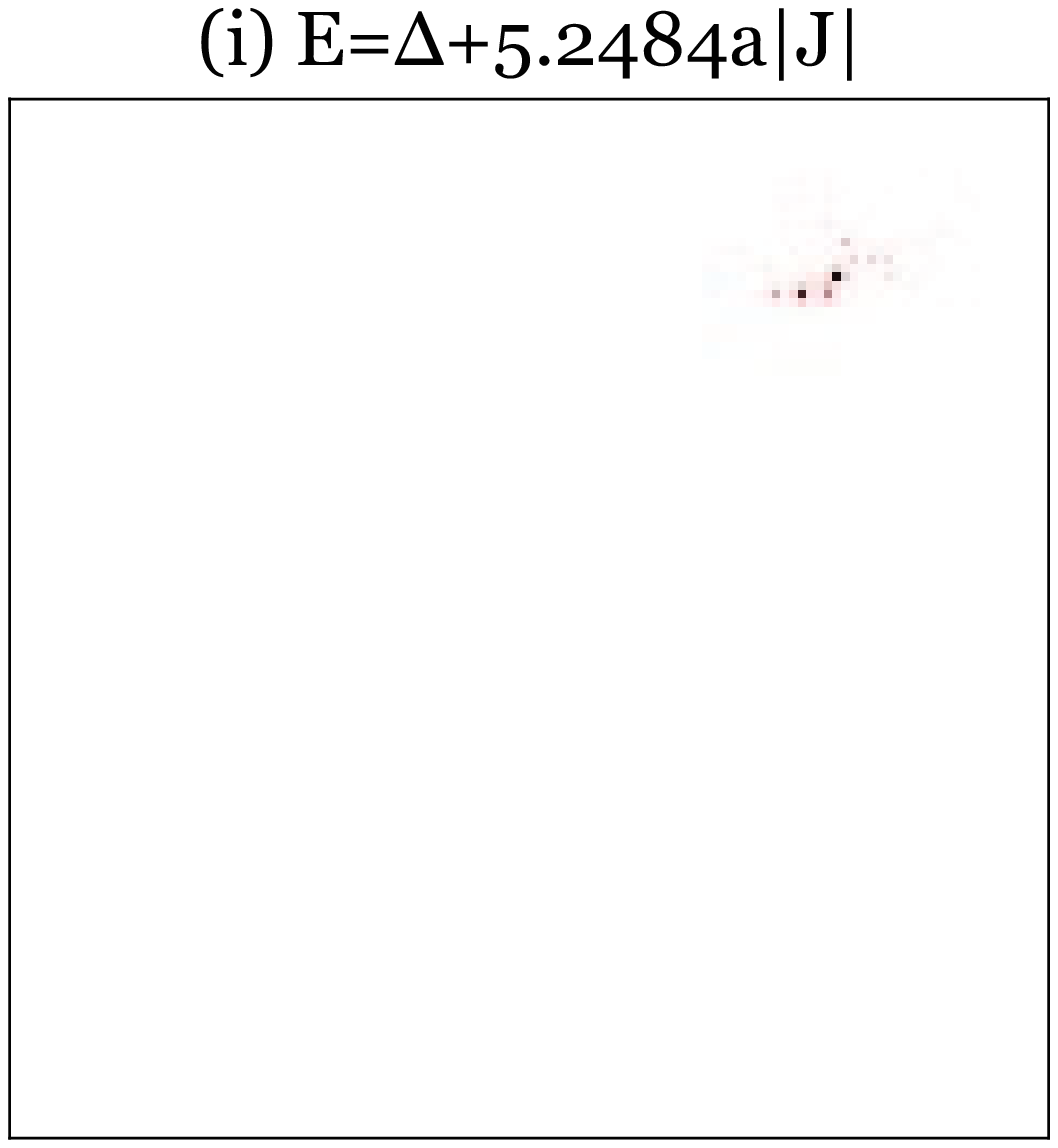}
	\includegraphics[scale=0.3]{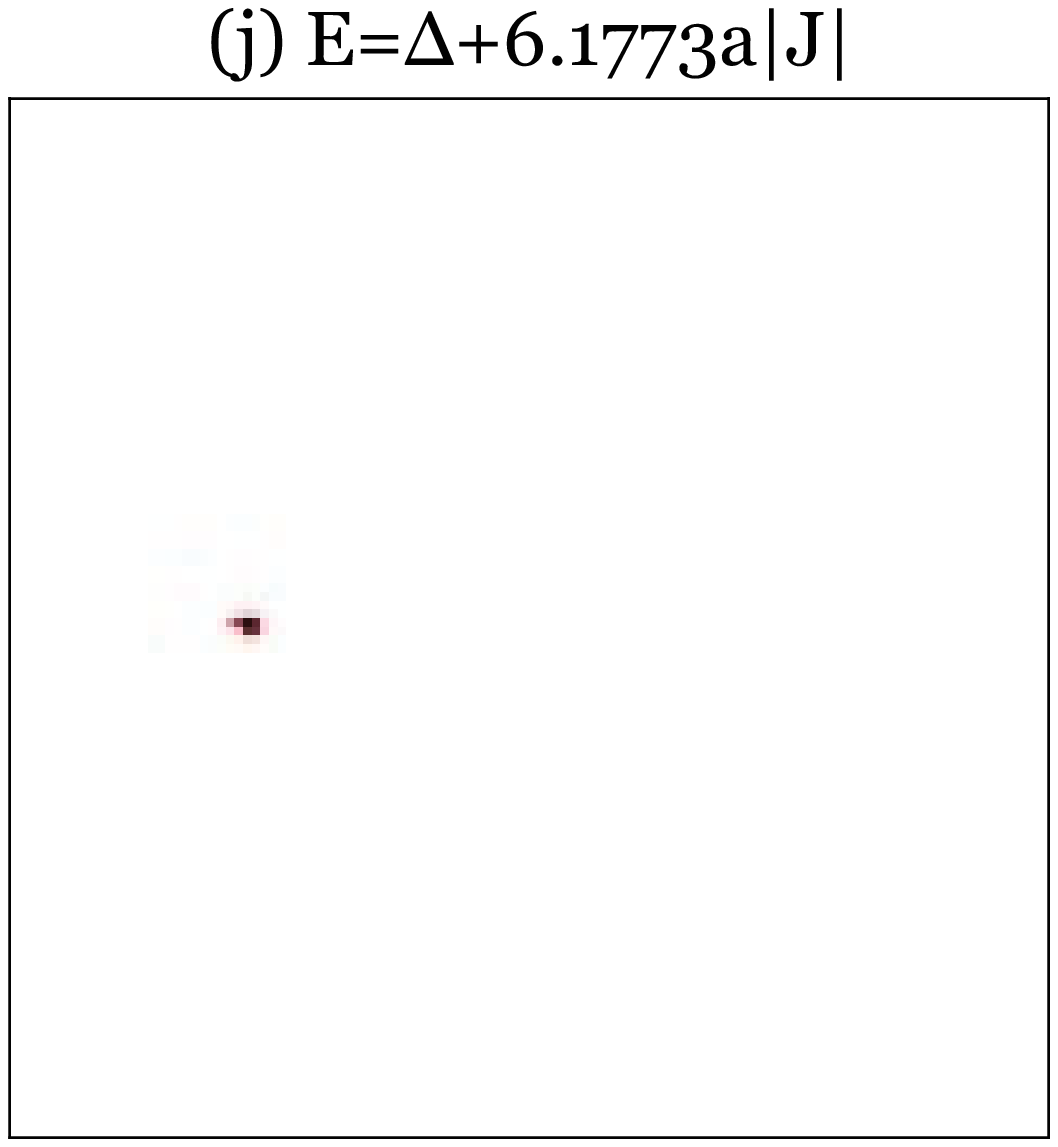}
	\includegraphics[scale=0.27]{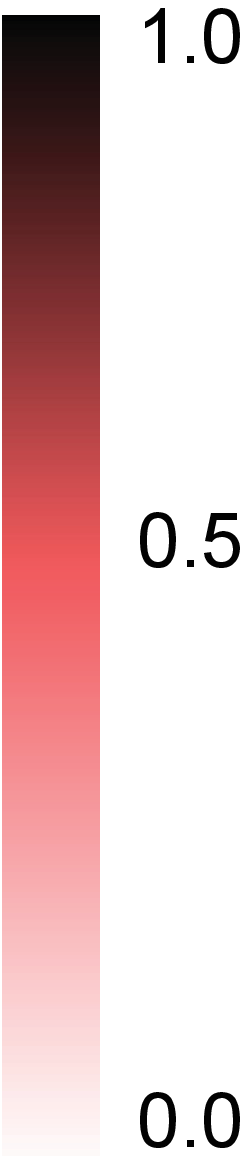}
  \hfil
  \caption{(Color online.) (a) Spatial distribution of defects with $c=0.1$, $u=3a|J|$, and $u_1=0$ in 2D system with the size $120\times120$ unit cells. (b)--(j) Numerically found color plots of wave functions amplitudes for the Hamiltonian of this disordered system which correspond to indicated eigenvalues $E$. Panels (b)--(e) give a picture of energy levels near the band bottom, panels (f)--(h) illustrate the band top, and panels (i) and (j) describe the impurity band corresponding to the localized level in the first order in $c$ (see also Fig.~\ref{dos2} for DOS found for the same parameters). All states in the impurity band are localized. $\Delta$ is the gap value for the particular disorder realization.}
  \label{loc2d}
\end{figure}

The analytical approach is not valid for states near the top and the bottom of the band due to localization of excitations that is illustrated by Fig.~\ref{loc2d} found numerically for a single disorder realization. We have also observed that IPR$\propto1/L^{\alpha d}$ for states inside the band far from its edges, where $\alpha<1$ (see Fig.~\ref{iprfig}(c) and \ref{iprfig}(d)). Although this behavior differs from that of ordinary propagating excitations ($1/L^d$), the localization length $\xi\propto L^\alpha$ is infinite in the thermodynamic limit in the considered 2D systems.

\begin{figure}
  \noindent
  \hfil
  \includegraphics[scale=0.45]{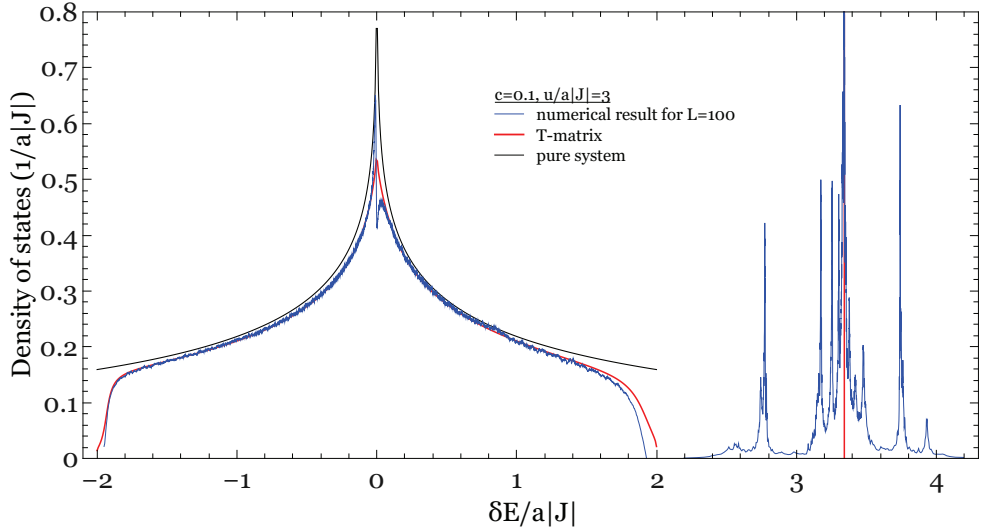}
  \hfil
  \caption{(Color online.) DOS of 2D systems with disorder in $\cal J$ or $\cal D$ only, where $\delta E=E-\Delta-2a|J|$, $c=0.1$, $u=3a|J|$, and $u_1=0$ (cf.\ Fig.~\ref{dos1}). Numerical results are obtained for the system size $100\times100$ unit cells.
	}
  \label{dos2}
\end{figure}

Defects impact on DOS is described by Eqs.~\eqref{ds1} and \eqref{ds2} which are difficult to treat analytically in 2D systems. As in 1D systems, Eq.~\eqref{ds2} has a solution at any finite $u$ outside the band, so that an isolated impurity level arises above and below the band for positive and negative $u$, respectively. The largest corrections to DOS inside the band appear near the bottom, the center and the top of the band which stem from singular derivatives in the numerator of the second term in Eq.~\eqref{ds1}. Due to these large corrections, the $T$-matrix approach does not work in these regions. These results are illustrated by Fig.~\ref{dos2} which demonstrates, in particular, our finding that in contrast to 1D systems there are no anomalies in spectrum corrections and DOS related to multiple-defects scattering processes (cf.\ Fig.~\ref{dos1}(b)). The numerical analysis of wave-functions shows that states around the anomaly at the band center remain propagating.

Taking into account disorder in $J_{ij}$ and performing calculations similar to those presented in Appendix~\ref{append} for 1D systems, we obtain corrections to quasiparticles energy and their damping which are cumbersome for arbitrary $\bf k$. However, these results turn out to be a simple modifications of Eqs.~\eqref{2dcorbot} in the vicinity of the spectrum minimum ($\kappa=|k-k_0|\ll1$):
\begin{eqnarray}
  E_{\bf k} &=& \Delta+\frac{a|J|}{2}\kappa^2+c \frac{\pi a |J|u'\left(\pi a|J|-u'\ln(\kappa/b_{21})+\left(b_{23} a\frac{J}{|J|}u_1+b_{24} u''\right)\right)}{\left(\pi a|J|-u'\ln(\kappa/b_{21})+\left(b_{23} a\frac{J}{|J|}u_1+b_{24} u''\right)\right)^2+\frac{\pi^2}{4}u^{\prime2}},\nonumber\\
  \gamma_{\bf k} &=& c \frac{\pi^2}{2} \frac{a |J| u^{\prime2}}{\left(\pi a|J|-u'\ln(\kappa/b_{21})+\left(b_{23} a\frac{J}{|J|}u_1+b_{24} u''\right)\right)^2+\frac{\pi^2}{4}u^{\prime2}},
	\label{spec2dbot}
\end{eqnarray}
where $u'=u-4au_1J/|J|-b_{22} au^2_1/|J|$, $u''=au^2_1/|J|$,
\begin{eqnarray}
  b_{22}&=&(5C_{20}-2C_{211}-C_{22}-8C_{21})/\pi=1.44, \nonumber \\
  b_{23}&=&C_{20}-C_{21}=1.57,  \\ b_{24}&=&C_{20}b_{22}-(C^2_{20}-2C_{211}C_{20}-C_{22}C_{20}-4C^2_{21})/\pi=4.79, \nonumber
\end{eqnarray}
where $C_{21}, C_{22}, C_{211}$ are model dependent coefficients as \eqref{C_20},
\begin{eqnarray}
  C_{21}&=&\pi a |J| \lim_{{\bf k}_1\rightarrow {\bf k}_0} \left(- \frac{1}{(2\pi)^2} \int_{\Omega} \frac{d^2k \cos k_x}{\varepsilon_{\bf k} - \varepsilon_{{\bf k}_1}} + \ln k_1\right), \nonumber\\
  C_{22}&=&-\pi a |J| \lim_{{\bf k}_1\rightarrow {\bf k}_0} \left( \frac{1}{(2\pi)^2} \int_{\Omega} \frac{d^2k \cos 2k_x}{\varepsilon_{\bf k} - \varepsilon_{{\bf k}_1}} + \ln k_1\right), \\
  C_{211}&=&-\pi a |J| \lim_{{\bf k}_1\rightarrow {\bf k}_0} \left( \frac{1}{(2\pi)^2} \int_{\Omega} \frac{d^2k \cos k_x \cos k_y}{\varepsilon_{\bf k} - \varepsilon_{{\bf k}_1}} + \ln k_1\right). \nonumber
\end{eqnarray}
We lead to the following expressions near the top of the band which resemble Eqs.~\eqref{2dcortop}:
\begin{eqnarray}
  E_{\bf k} &=& \Delta+4a|J|-\frac{a|J|}{2}\kappa^2 + c \frac{\pi a |J|u'\left(\pi a|J|+u'\ln(\kappa/b_{21})+\left(b_{23} a\frac{J}{|J|}u_1+b_{24} u''\right)\right)}{\left(\pi a|J|+u'\ln(\kappa/b_{21})+\left(b_{23} a\frac{J}{|J|}u_1+b_{24} u''\right)\right)^2+\frac{\pi^2}{4}u^{\prime2}},\nonumber\\
  \gamma_{\bf k} &=& c \frac{\pi^2}{2} \frac{a |J| u^{\prime2}}{\left(\pi a|J|+u'\ln(\kappa/b_{21})+\left(b_{23} a\frac{J}{|J|}u_1+b_{24} u''\right)\right)^2+\frac{\pi^2}{4}u^{\prime2}},
	\label{spec2dtop}
\end{eqnarray}
where now $u'=u+4au_1J/|J|+b_{22} au^2_1/|J|$ and $u''=au^2_1/|J|$. The weak dependence of corrections to the spectrum on momentum remains in the case of two types of disorder. It is seen from Eqs.~\eqref{spec2dbot} and \eqref{spec2dtop} that similar to 1D systems a mutual reduction of contributions from two sorts of disorder arises at $u|J| \approx 4au_1J+b_{22}au^2_1$ near the band bottom and at $u|J| \approx -4au_1J-b_{22}au^2_1$ near its top.

Analysis of DOS shows that similar to 1D systems the disorder in $J_{ij}$ only leads to one impurity level above the band and one impurity level below it if $u_1$ lies outside the interval $-2<u_1/J<0$ and there are no isolated impurity levels for $u_1$ lying inside this interval.

Similar to 1D systems, one leads to the same results \eqref{2dcorbot}--\eqref{val2d} and \eqref{spec2dbot}, \eqref{spec2dtop} using the general form of the spectrum \eqref{spec0} near its minimum (or maximum) because mainly small $\kap_p$ contribute to Green's functions $G_{mn}$ at small $\kap$. Model-dependent quantities in these expressions which depend on the form of the spectrum at $\kap_p\sim1$ are constants $b$. They are of the order of unity.

\subsection{3D systems}

\begin{figure}
  \noindent
  \includegraphics[scale=0.4]{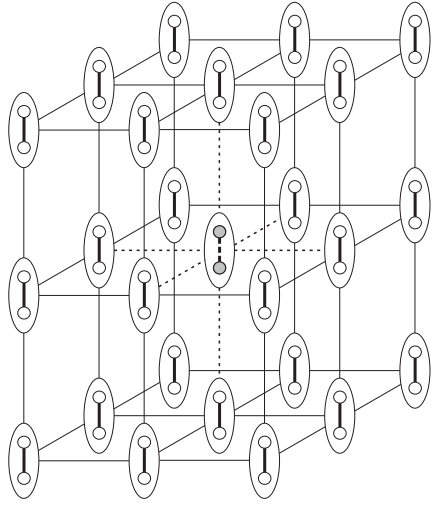}
  \hfil
  \caption{3D spin-$\frac12$ dimer system with imperfect bonds. Notations are the same as in Figs.~\ref{systems1D}(a) and \ref{SB1}.}
  \label{3D}
\end{figure}

3D spin-$\frac12$ dimer system under discussion is shown in Fig.~\ref{3D}. For the cubic lattice with interaction between nearest spins, the spectrum has the form
\begin{equation}
  \varepsilon_{\mathbf{k}} = \Delta + 3a|J| + aJ(\cos k_x + \cos k_y + \cos k_z),
	\label{spec03d}
\end{equation}
where $\Delta = 1-3a|J|$ at $H=0$. Green's function \eqref{Green1} can be represented as follows:
\begin{equation}
  G_{00}(E)=\frac{1}{\pi} \int^\pi_0 dz G^{(2D)}_{00}(E-a|J|-aJ \cos z),
  \label{Gr3D}
\end{equation}
where $G^{(2D)}_{00}$ is the Green's function \eqref{Green2D} for 2D systems. Eq.~\eqref{Gr3D} has the following form at $E=\varepsilon_{\mathbf{k}}$ near the spectrum minimum ($\kappa=|k-k_0|\ll1$):
\begin{equation}
   G_{00}(\varepsilon_{\mathbf{k}})\approx -\frac{1}{b_{31}a|J|} + i \frac{\kappa}{2 \pi a|J|},
	\label{g003d}
\end{equation}
where $b_{31}=1/C_{30}=2$, $C_{30}$ is a model dependent coefficient,
\begin{equation}
   C_{30}=\frac{a|J|}{(2\pi)^3} \int \frac{d^3k}{\varepsilon_{\bf k} - \varepsilon_{{\bf k}_0}}.
   \label{C_30}
\end{equation}
Using Eqs.~\eqref{Wk1}, \eqref{encor} and \eqref{g003d}, we obtain for the spectrum near its minimum in the case of disorder in $\cal J$ or $\cal D$ only
\begin{equation}
  E_{\bf k} = \Delta + \frac{a|J|}{2}\kappa^2 + c \frac{b_{31}a|J|u}{u + b_{31}a |J|},
	\qquad
	\gamma_{\bf k} = c \kappa \frac 1\pi \frac{b_{31}^2a|J|u^2}{\left( u + b_{31}a |J|\right)^2}.
	\label{spec3dbot}
\end{equation}
It is seen from Eqs.~\eqref{spec3dbot} that the quasiparticle energy acquires a small correction and $\gamma_{\bf k}\sim c\kappa$ if $|u+b_{31}a |J||\gg |u|\kappa$. However, the damping enhances greatly, $\gamma_{\bf k}\sim c/\kappa$, if $|u+b_{31}a |J||\ll |u|$ that signifies an appearance of a resonant scattering by defects in the first order in $c$.

In the vicinity of the spectrum maximum, one obtains the following results (cf.\ Eqs.~\eqref{spec3dbot}):
\begin{equation}
  E_{\bf k} = \Delta + 6 a |J| - a\frac{|J|}{2}\kappa^2 - c \frac{b_{31}a|J|u}{u-b_{31}a |J|},
	\qquad
	\gamma_{\bf k} = c \kappa \frac 1\pi \frac{b_{31}^2a|J|u^2}{\left( u-b_{31}a |J|\right)^2}.
	\label{spec3dtop}
\end{equation}
The resonant scattering takes place in this case if $|u-b_{31}a |J||\ll |u|$. If conditions $|u \pm b_{31}a |J||\ll |u|$ are not satisfied, the range of Eqs.~\eqref{spec3dbot} and \eqref{spec3dtop} validity is given by inequality $\kappa \gg c$.

\begin{figure}
  \noindent
  \hfil
  \includegraphics[scale=0.4]{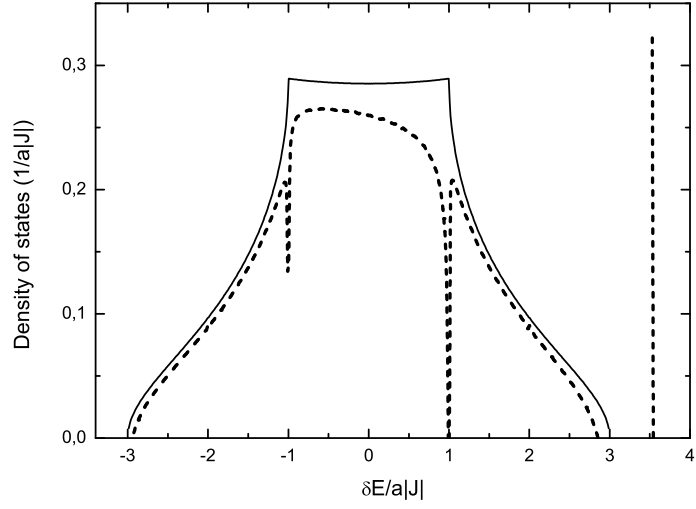}
  \hfil
  \caption{DOS of 3D systems, where $\delta E=E-\Delta-3a|J|$, $c=0.1$, $u=3a|J|$, and $u_1=0$. Solid and dashed lines are for pure and discorded systems, respectively.}
  \label{dos3}
\end{figure}

Similar to 2D systems, Eqs.~\eqref{g003d}--\eqref{spec3dtop} are valid in other gapped models in which the spectrum differs from \eqref{spec03d} but depends quadratically on the momentum near its minimum and maximum. The model dependent constant $b_{31}$ is of the order of unity in this case.

Effect of defects on DOS is illustrated by Fig.~\ref{dos3}. At $|u|<2a|J|$, there are no solutions of Eq.~\eqref{ds2} and there are no isolated impurity levels outside the band. If $|u|$ is large enough, $|u|>2a|J|$, the system has a localized level above or below the band for $u>0$ and $u<0$, respectively. Large corrections to DOS inside the band appear near its top and the bottom as well as at $E=\Delta+3a|J|\pm a|J|$ (see Fig.~\ref{dos3}) which stem from derivatives in the numerator of the second term in Eq.~\eqref{ds1}. The results obtained in the first order in $c$ are not valid near these anomalies.

Taking into account the disorder in $J_{ij}$, one obtains for the spectrum near the band bottom (cf.\ Eqs.~\eqref{spec3dbot})
\begin{eqnarray}
	\label{spec3dbot2}
  E_{\bf k} &=&\Delta+\frac{a|J|}{2}\kappa^2+cu' a|J|,
\quad
  \gamma_{\bf k}=c\kappa\frac{u^{\prime2}}{2\pi}a|J|,\\
\mbox{where }	u' &=& b_{31}\frac{u-6au_1J/|J|-b_{32} au^2_1/|J|}{u+b_{31}a|J|-b_{33}au_1J/|J|-b_{34}au^2_1/|J|},\nonumber
\end{eqnarray}
\begin{eqnarray}
  b_{32}&=&(21 C_{30}+3 C_{32}+12 C_{311}-36C_{31}) /2=3, \nonumber \\
  b_{33}&=&6 C_{31}/C_{30}=2, \\
  b_{34}&=&3(C^2_{30}+4C_{311}C_{30}+C_{30}C_{32}-6 C^2_{31})=1, \nonumber
\end{eqnarray}
where $C_{31},C_{32},C_{311}$ are model dependent constants as \eqref{C_30},
\begin{eqnarray}
  C_{31}&=&\frac{a|J|}{(2\pi)^3} \int \frac{d^3k \cos k_x}{\varepsilon_{{\bf k}_0} - \varepsilon_{\bf k}}, \nonumber\\
  C_{32}&=&\frac{a|J|}{(2\pi)^3} \int \frac{d^3k \cos 2k_x}{\varepsilon_{\bf k} - \varepsilon_{{\bf k}_0}}, \\
  C_{311}&=&\frac{a|J|}{(2\pi)^3} \int \frac{d^3k \cos k_x \cos k_y}{\varepsilon_{\bf k} - \varepsilon_{{\bf k}_0}}. \nonumber
\end{eqnarray}

We have near the spectrum maximum (cf.\ Eqs.~\eqref{spec3dtop})
\begin{eqnarray}
		\label{spec3dtop2}
  E_{\bf k} &=& \Delta+6a|J|-\frac{a|J|}{2}\kappa^2 - cu'a|J|,
\quad
  \gamma_{\bf k}=c\kappa\frac{u^{\prime2}}{2\pi}a|J|,\\
\mbox{where }	u' &=& b_{31}\frac{u+6au_1J/|J|+b_{32} au^2_1/|J|}{u-b_{31}a|J|+b_{33}au_1J/|J|+b_{34}au^2_1/|J|}.\nonumber
\end{eqnarray}
Similar to lower dimensions considered above, the phenomenon of corrections compensation
from two types of disorder arises in 3D systems as well: all corrections vanish at $u|J|=6au_1J+3au^2_1$ and $u|J|=-6au_1J-3au^2_1$ near the spectrum minimum and maximum, respectively.

For disorder in $J_{ij}$ only, analysis of DOS shows that similar to 1D and 2D systems one impurity level above the band and one impurity level below it appear if $u_1$ lies outside the interval $-2.75<u_1/J<0.75$ and there are no isolated impurity levels for $u_1$ lying inside this interval that is wider in 3D systems compared to 1D and 2D ones.

\section{Summary and conclusion}
\label{sum}

To summarize, we develop a theory based on the $T$-matrix approach which describes gapped phases in 1D, 2D, and 3D spin systems with bond disorder and with weakly interacting bosonic elementary excitations. Low-field paramagnetic and high-field fully saturated phases in dimerized spin-$\frac12$ magnets and integer-spin systems with large single-ion easy-plane anisotropy are considered in detail as examples. We discuss two sorts of disorder: i) that in intradimer coupling constants $\cal J$ or in the value of one-ion anisotropy $\cal D$ and ii) disorder in small exchange coupling constants $J_{ij}$ between spins from different dimers or spins on neighboring sites (in large-$\cal D$ systems).

For disorder in $\cal J$ or $\cal D$ only, we derive in the first order in the defects concentration $c$ the following expressions for corrections to propagating excitations energies and their damping: Eqs.~\eqref{spec1d} for 1D systems, Eqs.~\eqref{2dcorbot} and \eqref{2dcortop} for 2D systems, and Eqs.~\eqref{spec3dbot} and \eqref{spec3dtop} for 3D ones. It is found that the analytical approach does not work for states near the band edges so that ranges of the analytical results validity are given by Eqs.~\eqref{val1d} in 1D systems and by Eqs.~\eqref{val2d} in 2D and 3D ones. We demonstrate by performing numerical calculations that imaginary parts of the Green's function $\chi''({\bf k},\omega)$ show non-Lorentzian peaks at momenta for which the analytical approach does not work. Analysis of the corresponding wave functions demonstrates the localized nature of states in the band near its edges (see Fig.~\ref{loc2d} for the 2D system). Other states in the band remains propagating in 2D systems (and the same result is expected for 3D ones). In contrast, all states in the band turn out to be localized in 1D bosonic systems that resembles the situation in 1D electronic systems.  Besides, we find numerically that the analytical approach does not work in 1D systems if $c|u/aJ|\agt1$ due to multiple-defects resonance scattering that leads to anomalies in corrections to the spectrum and DOS (see Fig.~\ref{dos1}(b)). Analytical consideration of DOS shows that a localized impurity level arises above and below the band for any positive and negative $u$, respectively, in 1D and 2D systems whereas only $|u|>2a|J|$ leads to the isolated level in 3D systems.

Taking into account also the disorder in $J_{ij}$, we obtain in 1D systems for the spectrum and the damping Eqs.~\eqref{spec1d22}. Eqs.~\eqref{spec2dbot} and \eqref{spec3dbot2} give the spectrum and the damping in 2D and 3D systems, respectively, near the spectrum minimum, whereas Eqs.~\eqref{spec2dtop} and Eqs.~\eqref{spec3dtop2} are corresponding expressions in the vicinity of the spectrum maximum. In all dimensions, we find a phenomenon of mutual reduction of corrections to the spectrum and the damping from two types of disorder when certain relations are fulfilled involving $u$ and $u_1$. For disorder in $J_{ij}$ only, analytical results for DOS show that one impurity level above the band and one impurity level below it appear if $u_1$ lies outside the interval $-2<u_1/J<0$ in 1D and 2D systems and outside the interval $-2.75<u_1/J<0.75$ in 3D systems. There are no isolated impurity levels for $u_1$ lying inside these intervals.

Notice that expressions for the spectrum of propagating modes should also work at small temperature in the vicinity of $H_{bg1}$ or $H_{bg2}$ (see Fig.~\ref{th}). If there are no impurity levels inside the gap, the gap value can be reduced to zero by magnetic field. As a result the ratio of the long-wavelength quasiparticle damping to its energy can reach the value of $c/k^2$ (for 2D systems) in a wide range of parameters. Although this ratio is much smaller than unity in the range of this result validity $1\gg k\gg \sqrt c$ (as it must be for propagating excitations) it is much greater than $c$, the maximum value of $\gamma_{\bf k}/\varepsilon_{\bf k}$ obtained before for long-wavelength magnons in magnetically ordered magnets. \cite{wan,2dvac,syromyat1,syromyat2}

The results obtained can be relevant to other gapped phases in bond disordered spin systems both with and without a long range magnetic order. For instance, the phenomenon of localization of states near the band edges was observed theoretically in ferromagnets with random easy-axis anisotropy. \cite{medved} Eqs.~\eqref{spec1d22}, \eqref{spec2dbot}, \eqref{spec2dtop}, \eqref{spec3dbot2}, and \eqref{spec3dtop2} are derived using the general form of the spectrum \eqref{spec0} near its minimum (maximum) and using the general form of the impurity operators \eqref{v1} and \eqref{v2}. Then, they can be used for analysis of other gaped phases in other systems.

Results of recent neutron measurements of quasiparticles spectra at $H<H_{c1}$ in bond disordered dimer systems IPA-Cu(Cl$_x$Br$_{1-x}$)$_3$ (Ref.~\cite{nafradi}) and $\rm (C_4H_{12}N_2)Cu_2(Cl$$_{1-x}$Br$_x$)$_6$ (Ref.~\cite{huv2}) were interpreted under assumption that all excitations in the band are conventional wavepackets. As we see above for 1D systems, localized state can behave as a short-wavelength wavepacket. However such a behavior observed experimentally for states lying near the band bottom (corresponding to long-wavelength quasiparticles in pure systems) is quite puzzling. Our results demonstrate pronounced non-Lorentzian shape of Green's function imaginary part for states near the band bottom. Even according to the general theorem \cite{theorem} such states should be localized in these materials because no impurity levels arise in the gap (see Introduction). This point needs further experimental and theoretical analysis. Another point we leave for future studies is the influence of the quasiparticle interaction at low-field phases. This interaction is expected to play important role in real systems in which the gap value at $H=0$ is of the order of the band width.

\begin{acknowledgments}

This work is supported by RSF grant No.\ 14-22-00281, RFBR Grants No.\ 12-02-01234 and No.\ 12-02-00498. A.V.\ Sizanov acknowledges Saint-Petersburg State University for research grant 11.50.1599.2013.

\end{acknowledgments}

\appendix

\section{Disordered 1D systems}
\label{append}

In this appendix we provide some details of our consideration of the dimer spin ladder with imperfect intra- and interdimer coupling (see Fig.~\ref{systems1D}(a)) and of the integer spin chain with imperfect single-ion easy plane anisotropy and exchange coupling (see Fig.~\ref{systems1D}(b)).

The matrix form of the perturbation given by a sum of Eqs.~\eqref{v1} and \eqref{v2} reads in the one-particle basis $|0\rangle$, $|1\rangle$, and $|2\rangle$, where $|i\rangle$ denotes the state with one particle on $i$-th rung or site (see Fig.~\ref{systems1D}(a)),
\begin{equation}
  V_{nm}=\left[
           \begin{array}{ccc}
             0 & a u_1/2 & 0 \\
             a u_1/2 & u & a u_1/2 \\
             0 & a u_1/2 & 0 \\
           \end{array}
         \right].
\end{equation}
Further analysis is simplified by using the basis of irreducible representations of the symmetry point group:
$
  |\alpha,R(\alpha) \rangle = \sum^2_{(i=0)} U(i,\alpha,R(\alpha)) |i \rangle,
$
where $\alpha, R(\alpha)$ denotes basis wave-functions of irreducible representation $\alpha$. As reflection is the only nontrivial symmetry element of the system, new wave functions are either symmetric or antisymmetric and one has for them $|1\rangle$, $(|0\rangle+|2\rangle)/\sqrt{2}$, and $(|0\rangle-|2\rangle)/\sqrt{2}$. The corresponding matrices, which generate basis states for the representation, have the form
\begin{equation}
  T_s=\left[
           \begin{array}{cc}
             0 & 1/\sqrt{2}\\
             1 & 0 \\
             0 & 1/\sqrt{2} \\
           \end{array}
         \right],
	\qquad
	 T_p=\left[
           \begin{array}{c}
             1/\sqrt{2} \\
             0 \\
             -1/\sqrt{2} \\
           \end{array}
         \right].
\end{equation}

Corrections to quasiparticles spectra are defined by $T(k,E)$, which reads in this case as
\begin{equation}
  T(k,E)=\sum_{\mu=s,p} \psi^+(k) T_\mu (T^+_\mu V T_\mu) (T^+_\mu [I-G(E)V]^{-1} T_\mu)T^+_\mu \psi(k),
	\label{w1d}
\end{equation}
where $I$ is the identity matrix,
\begin{equation}
  \psi(k)=\left[
    \begin{array}{c}
      e^{-ik} \\
      1\\
      e^{ik}
    \end{array}
  \right],
\end{equation}
and elements of the Green's function matrix $G_{nm}$ \eqref{Green1} depend only on $|n-m|$
\begin{equation}
  G_{nm}=\left[
    \begin{array} {ccc}
      G_0 & G_1 & G_2 \\
      G_1 & G_0 & G_1 \\
      G_2 & G_1 & G_0 \\
    \end{array}
  \right],
\end{equation}
where $G_0$ is given by Eq.~\eqref{g001d} and
\begin{equation}
\label{g1g2}
  G_1(E) = \frac{1}{N}\sum_k \frac{e^{i k}}{E-\varepsilon_{k}-i 0},
	\qquad
  G_2(E) = \frac{1}{N}\sum_k \frac{e^{i 2 k}}{E-\varepsilon_{k}-i 0}.
\end{equation}
The contribution from antisymmetric representation $p$ is equal to zero in Eq.~\eqref{w1d} and the symmetric one gives
\begin{equation}
  T(k,E)=\frac{1}{D_s(E)} \left( u + \frac{a^2 u^2_1(G_0(E)+G_2(E))}{2} + 2 a u_1(1-a u_1 G_1(E)) \cos k + a^2 u^2_1 G_0(E) \cos^2 k\right),
	\label{w12}
\end{equation}
where
\begin{equation}
  D_s(E)=1-u G_0(E)-2 a u_1 G_1(E)+a^2 u^2_1 G^2_1(E)-\frac{a^2 u^2_1 G^2_0(E)}{2}-\frac{a^2 u^2_1 G_0(E) G_2(E)}{2}.
\end{equation}
One obtains from Eqs.~\eqref{g1g2} after simple calculations
\begin{equation}
  G_1(E)= \begin{cases}
    \frac{1}{aJ}\left(\frac{E-\Delta-a|J|}{\sqrt{(E-\Delta-a|J|)^2-a^2 J^2}}-1\right), & E>\Delta+2a|J|,\\
    \frac{1}{aJ}\left(i\frac{E-\Delta-a|J|}{\sqrt{a^2 J^2-(E-\Delta-a|J|)^2}}-1\right),& \Delta<E<\Delta+2a|J|, \\
    -\frac{1}{aJ}\left(\frac{E-\Delta-a|J|}{\sqrt{(E-\Delta-a|J|)^2-a^2 J^2}}+1\right), & E<\Delta,
  \end{cases}
\end{equation}
\begin{equation}
  G_2(E)= \begin{cases}
    \frac{2(E-\Delta-a|J|)^2}{a^2J^2\sqrt{(E-\Delta-a|J|)^2-a^2 J^2}}-\frac{1}{\sqrt{(E-\Delta-a|J|)^2-a^2 J^2}}-\frac{2(E-\Delta-a|J|)}{a^2J^2}, & E>\Delta+2a|J|, \\
    i\frac{2(E-\Delta-a|J|)^2}{a^2J^2\sqrt{a^2 J^2-(E-\Delta-a|J|)^2}}-i\frac{1}{\sqrt{a^2 J^2-(E-\Delta-a|J|)^2}}-\frac{2(E-\Delta-a|J|)}{a^2J^2}, & \Delta<E<\Delta+2a|J|, \\
    -\frac{2(E-\Delta-a|J|)^2}{a^2J^2\sqrt{(E-\Delta-a|J|)^2-a^2 J^2}}+\frac{1}{\sqrt{(E-\Delta-a|J|)^2-a^2 J^2}}-\frac{2(E-\Delta-a|J|)}{a^2J^2}, & E<\Delta.
  \end{cases}
\end{equation}
At $E=\varepsilon_{k}$, Eqs.~\eqref{g001d} and \eqref{g1g2} give
\begin{eqnarray}
\label{g0}
    G_0(\varepsilon_{k}) &=& i \pi g_0(\varepsilon_{k}),\\
    G_1(\varepsilon_{k}) &=& -\frac{1}{a J} + i \pi g_0(\varepsilon_{k}) \cos k,\\
    G_2(\varepsilon_{k}) &=& i \pi g_0(\varepsilon_{k})\cos{(2k)} - \frac{2 \cos k}{a J},
		\label{g2}
\end{eqnarray}
where $g_0(\varepsilon_{k})=1/(\pi|aJ\sin k|)$ is the pure system DOS and we get from Eq.~\eqref{w12}
\begin{equation}
  T(k,\varepsilon_{k})=\frac{(u+2 a u_1 \cos k+a u^2_1 \cos k /J)}{(1+ u_1/J)^2-i \pi g_0(\varepsilon_{k})(u+2 a u_1\cos k +a u^2_1 \cos k/J)}.
	\label{w1d2}
\end{equation}
One leads to Eqs.~\eqref{spec1d22} from Eqs.~\eqref{w1d2} and \eqref{encor}.

We turn now to DOS the general expression for which
\begin{equation}
  g(E)=g_0(E)+\frac{c}{\pi}\Im\left(\frac{d}{dE} \ln Det|1-G^0 V|\right)
\end{equation}
can be rewritten in the following form using the transformation to the irreducible representations basis:
\begin{equation}
  g(E)=g_0(E)+\frac{c}{\pi} \sum_{\mu=s,p} \frac{D'_\mu(E)}{D_\mu(E)},
	\label{g1d2}
\end{equation}
where the prime denotes the derivative on $E$. The contribution from $\mu=p$ is equal to zero in Eq.~\eqref{g1d2} and we obtain
\begin{equation}
  g(E) = g_0(E)+\frac{c}{\pi}\frac{\Re(D_s(E)(\Im (D_s(E)))'-(\Re(D_s(E)))'\Im(D_s(E))}{(\Re(D_s(E)))^2+(\Im(D_s(E)))^2}.
\end{equation}
Roots of equation $\Re(D_s(E))=0$ can give locations of virtual resonance levels inside the band and positions of isolated impurity levels outside the band. Using Eqs.~\eqref{g0}--\eqref{g2}, we lead after tedious transformations to the following quadratic equation on $x=(E-\Delta-a|J|)/a|J|$:
\begin{equation}
  (1+4t_1+2t^2_1)x^2-2t_0t_1(2+t_1)x-(t^2_0+(1+t_1)^4)=0,
	\label{eqx}
\end{equation}
where $t_0=u/a|J|$ and $t_1=u_1/J$. Solutions of Eq.~\eqref{eqx}
\begin{equation}
  x=\frac{t_0t_1(2+t_1) \pm (1+t_1)^2 \sqrt{t^2_0+1+4t_1+2t^2_1}}{1+4t_1+2t^2_1}
\end{equation}
determine location of DOS peculiarities and they should also satisfy the following condition:
\begin{equation}
  \frac{t_0}{x}-1+(1+t_1)^2\geq0.
\end{equation}
For $u_1=0$, one obtains Eq.~\eqref{imp1d} for the energy of the isolated level that is modified as follows at $|u|\gg a|u_1|$:
\begin{equation}
	E_d = \Delta+a|J|+{\rm sign}(u) \sqrt{a^2J^2+u^2} \left(1-2\frac{u_1}{J}\right)+2u\frac{u_1}{J}.
\end{equation}
When $u=0$, solutions exist at $t_1>0$ or $t_1<-2$ only. DOS peculiarities lie outside the band and we have for energies of two isolated impurity levels arising above and below the band
\begin{equation}
  E_d = \Delta+a|J| \pm a|J|\frac{(1+t_1)^2}{\sqrt{1+4t_1+2t^2_1}}.
\end{equation}

\bibliography{bibliography}

\begin{thebibliography}{28}
\bibitem[{foo()}]{foot}
\bibinfo{howpublished}{One expects that $H_{c1}$ and $H_{c2}$ in
  Fig.~\ref{th}(b) do not differ significantly from the critical field values
  of the pure system if the defect concentration $c$ is small.}
\end{thebibliography}

\end{document}